%% file: DGHpaper.tex
\documentclass[a4paper,12pt]{article}
\pdfoutput=1

\usepackage{amsmath,amsfonts,amssymb,epsfig}
\usepackage{slashed}
\numberwithin{equation}{section}

\usepackage{graphicx}
\usepackage{cancel}
\usepackage{cite}
\usepackage{color}

\usepackage{xspace}

\newcommand{\SARAH}{{\tt SARAH}\xspace}
\newcommand{\SPheno}{{\tt SPheno}\xspace}

\def\pdir{Paperplots/}

\def\gev{\ \mathrm{GeV}}
\def\gevsq{\ (\mathrm{GeV})^2}

\newcommand{\exclude}[1]{}
\def\nn{\nonumber}

\def\beq{\begin{equation}}
\def\eeq{\end{equation}}
\def\bal{\begin{align}}
\def\eal{\end{align}}

\def\Br{\mathrm{Br}}

\def\2b2[#1,#2][#3,#4]{\left( \begin{array}{cc} #1 & #2 \\ #3 & #4 \end{array}
\right)}
\def\3b3[#1,#2,#3][#4,#5,#6][#7,#8,#9]{\left( \begin{array}{ccc} #1 & #2 &#3 \\
#4 & #5 & #6\\#7&#8&#9\end{array} \right)}
\def\thv[#1,#2,#3]{\left( \begin{array}{c} #1 \\ #2 \\ #3 \end{array} \right)}
\def\twv[#1,#2]{\left( \begin{array}{c} #1 \\ #2 \end{array} \right)}

\newcommand{\C}[1]{\mathcal{#1}}

\author{Karim Benakli$^1$\footnote{kbenakli@lpthe.jussieu.fr},
Mark~D.~Goodsell$^2$\footnote{mark.goodsell@cern.ch, mark.goodsell@cpht.polytechnique.fr}
and Florian Staub$^3$\footnote{fnstaub@th.physik.uni-bonn.de}
}
\date{}
\title{\vspace{-3cm}\small{
\hfill{BONN-TH-2012-22}}\\[2cm]
\huge{Dirac Gauginos and the 125 GeV Higgs}
}
\begin{document}
\maketitle
\vspace{-1cm}
\begin{center}
\emph{$^1$Laboratoire de Physique Theorique et Hautes Energies, CNRS, \\
UPMC Univ Paris VI Boite 126, 4\\Place Jussieu, 75252 Paris cedex 05, France
} \\[3mm]
\emph{$^2$CPhT, Ecole Polytechnique, 91128 Palaiseau, France} \\[3mm]
\emph{$^3$Bethe Center for Theoretical Physics \& Physikalisches Institut der Universit\"at Bonn\\Nu{\ss}allee 12, 53115 Bonn, Germany}
\end{center}
\abstract{We investigate the mass, production and  branching ratios of a 125 GeV Higgs in
models with Dirac gaugino masses. We give a discussion of naturalness, and describe how deviations from the Standard Model in
the key Higgs search channels can be simply obtained. We then perform parameter scans using 
 a \SARAH package upgrade, which produces \SPheno code that calculates all relevant quantities, including electroweak precision
and flavour constraint data, to a level of accuracy previously impossible for this class of models. We study three different variations on the
minimal Dirac gaugino extension of the (N)MSSM.  
}

\section{Introduction}

The LHC experiments have claimed the discovery of a new particle at 125 $\pm$
1 GeV \cite{Atlas:2012gk,CMS:2012gu}. Its production and decays make it a good candidate for
the Higgs boson. However, more experimental data is needed for a precise
determination of its quantum numbers and interactions to know
if the Higgs sector is the Standard Model one, or an extended version, for
instance as required in supersymmetric models. A clear indication of 
non-minimality would be a significant excess or deficit in at least one decay channel, but the Higgs mass
itself can also be regarded as such, since it is somewhat high for the most natural version of the MSSM. This 
is one motivation for this work, in which we study
the main properties of such a Higgs boson in  models with
Dirac gaugino masses \cite{Fayet:1978qc,Polchinski:1982an,Hall:1990hq,Fox:2002bu,Nelson:2002ca,Antoniadis:2005em,Antoniadis:2006uj,Hsieh:2007wq,Amigo:2008rc,Benakli:2008pg,Belanger:2009wf,Benakli:2009mk,Chun:2009zx,Benakli:2010gi,Carpenter:2010as,Kribs:2010md,Chun:2010hz,Choi:2010an,Benakli:2011kz,Abel:2011dc,Kumar:2011np,Benakli:2011vb,Davies:2011mp,Davies:2011js,Heikinheimo:2011fk,Rehermann:2011ax,Kribs:2012gx,Davies:2012vu,Argurio:2012cd}. These have numerous virtues compared to their Majorana counterparts, not least that they allow for increased naturalness \cite{Jack:1999ud,Jack:1999fa,Fox:2002bu,Benakli:2011kz,Goodsell:2012fm} which is particularly important in the light of recent LHC SUSY searches \cite{Brust:2011tb,Papucci:2011wy}; but also that the direct production of gluinos is suppressed, loosening the bounds from direct LHC searches \cite{Heikinheimo:2011fk,Kribs:2012gx}; they allow relaxation of flavour constraints such as due to $\mathrm{Br} (B \rightarrow s \gamma)$ \cite{Kribs:2007ac}; they preserve R-symmetry (allowing for simpler supersymmetry-breaking models) and can be motivated from higher-dimensional theories as being derived from an $N=2$ supersymmetry in the gauge sector at some scale. 

To give gauginos Dirac masses, we must add new chiral superfields in the
adjoint representation of each gauge group: a singlet $\mathbf{S}  =  S + \sqrt{2} \theta \chi_S + \cdots$,
an $SU(2)$  triplet $\mathbf{T} = \sum_{a=1,2,3}
\mathbf{T}^{(a)} = T^{(a)} + \sqrt{2} \theta \chi_T^{(a)} + \cdots $ and
an $SU(3)$  octet $\mathbf{O_g}  = \sum_{a=1 .. 8} O^{(a)}  + \sqrt{2} \theta \chi_O^{(a)}+ \cdots$. 
Of these, the triplet and singlet may have new renormalisable couplings with the Higgs 
which allow more possibilities to obtain the desired mass
range than in the MSSM. More precisely, the lightest Higgs mass in these models is
determined by opposite competing effects. On one hand, the presence of new
couplings in the extended scalar sector leads to  an enhancement of this mass by
allowing new contributions to the quartic Higgs coupling. On the other hand, the supersoft 
operators that include the Dirac mass induce new D-term couplings, which increase Higgs mixing
and thus tend to reduce the lightest Higgs mass. However, this is only potentially problematic if the
triplet scalar soft mass is small; but we shall demonstrate in section \ref{SEC:NATURAL} - along with a general discussion of naturalness in Dirac gaugino models - that it may be naturally large enough to avoid this problem. Higgs 
mixing involving the singlet induced by a Dirac Bino mass, however, then presents an intriguing opportunity: it allows
the decays of the Higgs to $b$ quarks and $\tau$ leptons to be suppressed while preserving the rate to $W$s and $Z$s, and an enhancement of the diphoton rate. We give an explanation of this and a discussion of the Higgs production and decay rates in section \ref{SEC:DECAYS}.

Previous attempts to quantitatively study the Higgs sectors of Dirac gaugino models have been hampered by the lack of numerical tools to do so; until now only one-loop effective potential calculations of the Higgs mass have been possible. This is in contrast to the MSSM, where the leading corrections are known to    three-loop order. However, with an upgrade of the \SARAH package described in section \ref{SEC:SARAH}, it is now possible to study models with arbitrary gauge groups and matter content: it can generate \SPheno code which calculates all one-loop pole masses and tadpoles, which allows a much more accurate determination of the Higgs mass. We implement a minimal Dirac-gaugino extension of the (N)MSSM which lends itself to four particularly interesting sub-classes of models, which we review in section \ref{SEC:SETUP}. One such class of models is the ``MSSM in disguise,'' where all the new scalars are too heavy to observe or mix substantially with the Higgs; this is a good toy scenario to use to test of the code, and we do just that. 

The \SPheno code produced by \SARAH can also calculate the branching ratios and production cross-sections of the Higgs, as well as electroweak precision observables such as $\Delta \rho$ and flavour constraints such as $\mathrm{Br} (B \rightarrow s \gamma)$. We take full advantage of the latter in investigating the ``MSSM without $\mu$ term'' \cite{Nelson:2002ca} in section \ref{SEC:MULESS}, comparing the results for  $\Delta \rho$ with approximations given in \cite{Nelson:2002ca}. Unfortunately those constraints in addition to those on chargino masses and the Higgs mass yield that model problematic. However, we propose a new model, which we call ``dynamica $\mu$ models,'' in which the singlet obtains a substantial expectation value - we show in model scans in section \ref{SEC:RBREAKING} that this not only alleviates all of the problems of the MSSM without $\mu$ term, but also allows for Higgs mixing and thus interesting deviations in the Higgs production rates and branching ratios that may better fit the current data than the standard model.

\section{One model into four}
\label{SEC:SETUP}

Adding Dirac gaugino masses to the MSSM introduces many extra parameters: not just 
the Dirac masses themselves, but also new superpotential couplings and soft terms.
 By making certain assumptions about the new parameters we can then arrive at different 
limits of the model with different phenomenology; we shall consider four such limits. 

The most general renormalisable superpotential that we can write down is
\begin{eqnarray}
W &=& Y_u \hat{u} \hat{q} H_u - Y_d \hat{d} \hat{q} H_d - Y_e \hat{e} \hat{l} H_d  +  \mu \mathbf{H_u\!\cdot\! H_d }\nn\\
&&  + \lambda_S \mathbf{SH_u\!\cdot\!H_d}  + 2  \lambda_T \mathbf{H_d\!\cdot\! T H_u} \nn\\
&&+L \mathbf{S}  + \frac {M_S}{2}\mathbf{S}^2 + \frac{\kappa}{3}
\mathbf{S}^3 + M_T \textrm{tr}(\mathbf{TT}) + M_O \textrm{tr}(\mathbf{OO}) \nn\\
&&+W_2
\label{NewSuperPotential}
\end{eqnarray}
where
\begin{align}
W_{2} =  \lambda_{ST}
\mathbf{S}\textrm{tr}(\mathbf{TT}) +\lambda_{SO} \mathbf{S}\textrm{tr}(\mathbf{OO})
 + \frac{\kappa_O}{3} \textrm{tr}(\mathbf{OOO}).
\label{AdjointSuperpotential}\end{align}

The usual scalar soft terms are
\begin{align}
\label{potential4}
- \Delta\mathcal{L}^{\rm scalar\ soft}_{\rm MSSM} =& [ T_u \hat{u} \hat{q} H_u - T_d \hat{d} \hat{q} H_d - T_e \hat{e} \hat{l} H_d  + h.c. ]\nn\\
&+ m_{H_u}^2 |H_u|^2 +
m_{H_d}^2 |H_d|^2
 + [B_{\mu} H_u\cdot H_d + h.c. ]\nn\\
&+ \hat{q}^i (m_q^2)_i^j \hat{q}_j + \hat{u}^i (m_u^2)_i^j \hat{u}_j + \hat{d}^i (m_d^2)_i^j \hat{d}_j + \hat{l}^i (m_l^2)_i^j \hat{l}_j  + \hat{e}^i (m_e^2)_i^j \hat{e}_j  
\end{align}
and there are  soft terms involving the adjoint scalars
\begin{eqnarray}
- \Delta\mathcal{L}^{\rm scalar\ soft}_{\rm adjoints} &= &  (t_S S + h.c.) \nn\\
&&+ m_S^2  |S|^2 + \frac{1}{2} B_S
(S^2 + h.c.)  + 2 m_T^2 \textrm{tr}(T^\dagger T) + (B_T \textrm{tr}(T T)+ h.c.)
\nonumber \\  &&+ 
[A_S \lambda_S SH_u\cdot H_d +  2 A_T \lambda_T H_d \cdot T H_u +
\frac{1}{3} \kappa  A_{\kappa} S^3 + h.c.] \nonumber \\ &&+ 2 m_O^2 \textrm{tr}(O^\dagger O) 
+ (B_O \textrm{tr}(OO)+ h.c.)
\label{Lsoft-DGAdjoint}
\end{eqnarray}
with the definition $H_u\cdot H_d = H^+_uH^-_d - H^0_u H^0_d$. Similarly there are the A-terms for $W_2$.

One limit of the above model would be to allow all parameters, including Majorana gaugino masses, 
to be significant and non-vanishing. Such models may have virtues due to the extra Higgs
 couplings and extra gaugino states (the charginos could contribute, for example, 
to enhancing the Higgs-to-diphoton decay channel) but we shall leave the exploration of 
this to future work. Instead, we shall explore models where the gaugino masses are entirely Dirac,
 taking as motivation the possibility of preserving R-symmetry in some sector of the theory
 (but not in all: it must ultimately be broken in any case). Of particular interest to us are:
\begin{itemize}
\item \emph{MSSM in disguise}: here we shall allow a $\mu$-term, and assume that the only source of R-symmetry violation arises in the supersymmetry-breaking sector, but permit only a $B_\mu$ term. This assumption will be preserved by renormalisation group running and so it can be justified by high-energy boundary conditions; non-zero A-terms would lead to Majorana gaugino masses. Since it is the ``MSSM in disguise'' we shall take the scalar singlet and triplet to be very massive (several TeV). We perform some scans over models of this type in section \ref{SEC:SARAH}. 
\item \emph{MSSM without $\mu$ term} ($\slashed{\mu}\rm{SSM})$: this is the scenario of \cite{Nelson:2002ca}, similar to the MSSM in disguise but taking $\mu = 0$. Here we shall insist that the singlet vev is small, so the chargino mass must be generated by the coupling $\lambda_T$. We investigate this scenario in section \ref{SEC:MULESS}.
\item \emph{Dynamical $\mu$ models}: in this scenario, we again take $\mu=0$ but allow non-zero $B_S$, leading to a substantial non-zero expectation value for the singlet. In this way, the vev and the coupling $\lambda_S$ lead to an effective $\mu$-term. Models of this type are very natural and interesting from the point of view of Higgs mixing: we perform scans over these models in section \ref{SEC:RBREAKING}.
\item \emph{Dynamical $\mu$ and $B\mu$ models}: this is the scenario of \cite{Benakli:2011kz} where we allow a non-zero $\kappa$, breaking R-symmetry in the visible sector, but allowing $\mu$ and $B_\mu$ to be generated via a non-zero singlet vev - so we can set all R-symmetry-breaking parameters in the supersymmetry breaking sector to zero. It is somewhat similar to the NMSSM, but the Dirac masses lead to some interesting differences. Models of this type can be very natural, but we leave scans of their parameter space to future work.
\end{itemize}

We review the tree-level properties of the generic case in appendix \ref{APP:TREE}; see also \cite{Benakli:2011kz}. However, common to all of the above scenarios is the assumption that Dirac gaugino masses dominate over Majorana ones. For this to be natural, A-terms must also be small, so in our searches we shall set (unless otherwise stated)  $W_2 = L = M_S = M_T = M_O = A_S = A_\kappa = A_T = 0$. 

\subsection*{Some notation}

We now introduce some notation relevant for the following: we redefine the singlet and triplet scalars in terms of real components $S \equiv \frac{1}{\sqrt{2}} ( v_S + s_R + i s_I), T^0 \equiv \frac{1}{\sqrt{2}} ( v_T + T_R + i T_I) $ and have an ``effective $\mu$-term'' $\tilde{\mu} \equiv \mu + \frac{1}{\sqrt{2}} ( v_S \lambda_S + v_T \lambda_T) $. The expectation values $v_S, v_T$ are associated with new non-trivial potential minimisation conditions, which we give in equation (\ref{EQ:vsvt0}). The scalars $s_R, T_R$ mix with the Higgs fields, with a $4\times 4$ mass matrix given in equation (\ref{eq:MassScalar}); the mixing will be important in section \ref{SEC:RBREAKING}.

\section{Numerical Setup}
\label{SEC:SARAH}
\subsection{Implementation in \SARAH and \SPheno}
For our numerical studies we have extended the Mathematica package \SARAH  \cite{Staub:2008uz,Staub:2009bi,Staub:2010jh,Staub:2012pb} 
to support Dirac gauginos and implemented our model \footnote{The support of Dirac gauginos as well
 as the used model files will become public with the next major upgrade to version 3.2.0 of \SARAH.}
 Afterwards we used the possibility of \SARAH to create source code for \SPheno \cite{Porod:2003um,Porod:2011nf} 
for the given model. The obtained code was compiled with \SPheno 3.1.2 and provides 
a fully functional spectrum calculator: the entire mass spectrum is calculated at one-loop using
 the full dependence on the momentum of the external particle. For a detailed discussion of the 
calculation of the loop corrected mass spectrum using \SARAH and \SPheno for extensions 
of the MSSM we refer the interested reader to Ref.~\cite{Staub:2010ty}. In addition, the produced \SPheno
 version includes routines to calculate the decay widths and 
branching ratios for all SM superpartners and Higgs fields. In general, these are pure tree-level calculation. 
However, in the case of the Higgs particles the loop induced decays into two photons and two gluons are included. 
 This calculation is comparable to the analytical results given in sec.~\ref{SEC:DECAYS}, 
but also the NLO corrections due to quarks and squarks are added
 as given in Ref.~\cite{Spira:1995rr}. Similarly, for the Higgs decays into quarks the dominant QCD corrections
 due to the gluon are taken into account \cite{Djouadi:1996}. Finally, the \SPheno version for Dirac
 gauginos also calculates the observables $b \to s\gamma$ and $\Delta \rho$ with the same precision as 
done in the MSSM by \SPheno 3.1.1 including all contributions due to the new states present in the considered model.
 The details of these calculations can be found in Ref.~\cite{Porod:2011nf} and references therein. \\

Since there are four non-trivial vacuum minimisation conditions (given in equations (\ref{EQ:hmin}), (\ref{MA1}) and (\ref{EQ:vsvt0})) we must use these to eliminate four parameters. From a top-down perspective, it would be preferable to specify the soft masses $m_S, m_T$ as was done in \cite{Benakli:2011kz}, and derive from that $v_S, v_T$. However, the equations are non-linear in these, and so it is preferable to instead take $v_S, v_T, \tan \beta, \mu$ as input parameters in the code, and treat $m_S^2, m_T^2, m_{H_u}^2, m_{H_d}^2$ as output parameters (calculated including one-loop tadpoles). As further inputs for our studies we use for our studies the soft-breaking terms as well as the superpotential couplings at the SUSY scale: the standard model gauge
 and Yukawa couplings are calculated at $M_{SUSY}$ using two-loop standard model RGEs from $M_Z$ \cite{Arason:1991ic}.  

Finally we note that, in the absence of the theoretical calculation, the code cannot include the two-loop corrections to the Higgs mass involving (Dirac) gluino masses that can be important in the MSSM and NMSSM. In those cases, the Higgs mass is usually increased by $2$ to $4 \gev$. Throughout we shall be conservative and allow a variation of $\pm 4\gev$ for the mass of the Higgs in the scans, but we expect that typically the shift will be positive. 

\subsection{Comparison with effective potential}

It is possible to calculate an approximation for the Higgs mass via the effective potential technique. In the decoupling limit of large $B\mu$, this yields 
\begin{align}
m_h^2 \simeq&  M_Z^2 c^2_{2\beta} + \frac{v^2}{2}(\lambda_S^2+\lambda_T^2) s^2_{2\beta} \nn\\
&+ \frac{3}{2\pi^2} \frac{m_t^4}{v^2} \bigg[ \log \frac{ m_{\tilde{t}_1}m_{\tilde{t}_2}}{m_t^2} + \frac{\mu^2 \cot^2 \beta}{m_{\tilde{t}_1}m_{\tilde{t}_2}} \bigg( 1 - \frac{\mu^2 \cot^2 \beta}{12m_{\tilde{t}_1}m_{\tilde{t}_2}}\bigg) \bigg] \nn\\
&+ v^2 \bigg[ \lambda_1 c_\beta^4 + \lambda_2 s_\beta^4 + 2 (\lambda_3 + \lambda_4 + \lambda_5)  c_\beta^2 s_\beta^2 + 4 (\lambda_6 c_\beta^2 + \lambda_7 s_\beta^2) s_\beta c_\beta\bigg] \nn\\
&\overset{\tan \beta \rightarrow \infty}{\longrightarrow} M_Z^2  + \lambda_2 v^2 + \frac{3}{2\pi^2} \frac{m_t^4}{v^2}  \log \frac{ m_{\tilde{t}_1}m_{\tilde{t}_2}}{m_t^2} 
\label{EQ:EffHiggsMass}\end{align}
where the first line is the tree-level mass, the second line is the usual contribution from stops with $m_{\tilde{t}_{1,2}} $ the physical masses and  $A_t$ set to zero; the $\lambda_i$ are the coefficients of the terms in the most general CP-conserving gauge-invariant potential up to quartic order (e.g. $\lambda_2$ is the coefficient of the $|H_u|^4$ operator) about zero vev \cite{Haber:1993an}. 
This is a good approximation when the Higgs vev is small compared to the energy scales being integrated out; 
in this case, we shall consider integrating out the adjoint scalars, which is appropriate for the MSSM in disguise. We give the full expression for the coefficients $\lambda_i$ in appendix \ref{APP:EFFPOT}, but an interesting limit is to consider $B_S = B_T = 0$ and neglect $v, v_S$ and the Dirac masses $m_{Di}$; then the contribution from the singlet and triplet scalars is  \cite{Benakli:2011kz}
\begin{align}
32\pi^2 \lambda_2 \supset& 2 \lambda_S^4 \log \frac{m_S^2}{v^2} + (g_2^4 - 4g_2^2 \lambda_T^2 + 10 \lambda_T^4) \log \frac{m_T^2}{v^2} \nn\\
&+ \frac{4 \lambda_S^2 \lambda_T^2 }{m_S^2 - m_T^2}\bigg[ m_S^2 \log \frac{m_S^2}{v^2} - m_T^2 \log \frac{m_T^2}{v^2} - ( m_S^2 - m_T^2) \bigg] \nn\\
\overset{ m_T^2 \rightarrow m_S^2}{\longrightarrow}& \bigg( g_2^4 - 4g_2^2 \lambda_T^2 + 2 ( \lambda_S^4 - 2 \lambda_S^2 \lambda_T^2 + 5 \lambda_T^4)\bigg) \log \frac{m_S^2}{v^2}
\end{align}
where in the last line we show the limit that the scalar masses become equal.

\begin{figure}
\begin{center}
\includegraphics[]{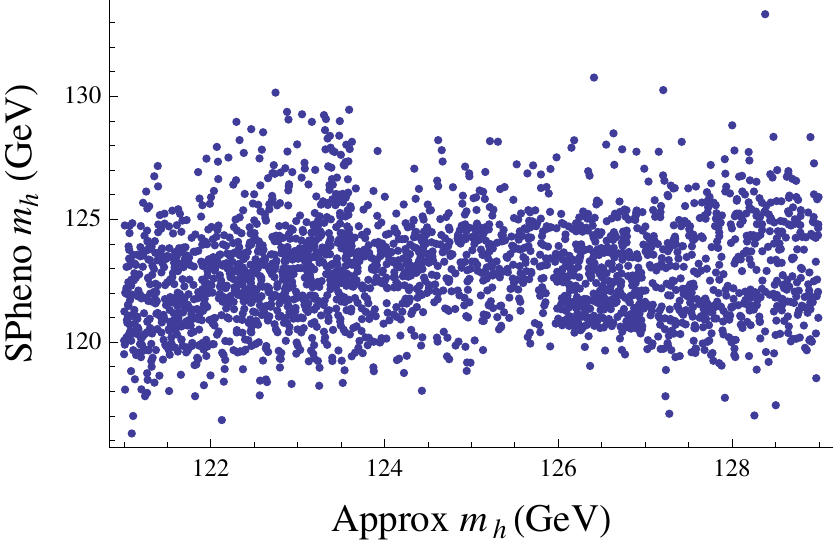}
\caption{Comparison of effective potential method (labelled ``approx $m_h$'') with a parameter scan using \SPheno code produced by \SARAH. Parameters are given in the text; a searched was performed using the \SPheno code, restricting to $m_h = 125 \pm 4 \gev$, and from the same input parameters for each model the one-loop Higgs mass is shown (note that this has a larger range). Only models with relatively low couplings $\lambda_S, \lambda_T $  (roughly $\lambda_S^2 + \lambda_T^2 < 1.2$).}
\end{center}
\label{FIG:CompareEffPotTan50}
\end{figure}

By performing a scan over parameters we can compare this approximation with the more accurate results 
from \SPheno using the code produced by \SARAH; we show the results with the choice of $\tan \beta = 50$, $m_{D2}= 600 \gev$, first two generation sfermion mass squareds of $4 \times 10^7 \gevsq$, third generation sfermion mass squareds $4 \times 10^6 \gevsq $  and scanning over $m_{D1} \in [-600,600 ] \gev, \mu \in [-750,750] \gev, B\mu \in [5000,10^6 ]\gevsq, \lambda_T \in [0, 1]$ while adjusting $\lambda_S$ to keep $m_h = 125 \pm 4 \gev$ in figure \ref{FIG:CompareEffPotTan50}. The expectation values $v_S, v_T$ were set by the tree level minimisation equation with $m_S^2, m_T^2 =2.5 \times 10^7 \gevsq$; this resulted in values close to this for $m_T^2$, while the resulting one-loop adjusted values for $m_S^2$ varied between $10^6$ and $10^{10} \gevsq$. As can be seen from the figure, there is good agreement between the two, although of course the approximate effective potential calculation exhibits a wider variation of masses; that there is no apparent correlation is unsurprising, essentially reflecting the error margin in the effective potential method.  

\section{Dirac gauginos and Natural SUSY}
\label{SEC:NATURAL}

One of the increasingly attractive features of Dirac gaugino models is that the supersoft operators allow for increased naturalness \cite{Jack:1999ud,Jack:1999fa,Fox:2002bu,Benakli:2011kz,Brust:2011tb,Papucci:2011wy,Goodsell:2012fm}; they do not appear in the RGEs for the soft masses and so they only affect the stop mass via a UV-finite correction, allowing for heavy gluinos with light stops. On the other hand, the Higgs potential is corrected at tree-level by two competing effects: one is the enhancement of the Higgs mass (at low $\tan \beta$) by the new couplings $\lambda_S, \lambda_T$, clearly evident in equation (\ref{EQ:EffHiggsMass}); the other is a reduction in the effective D-term Higgs quartic coupling due to the Dirac mass terms; if we ignore the superpotential couplings and integrate out the heavy scalars then  \cite{Benakli:2011kz,Fox:2002bu}
\begin{align}
\frac{g_Y^2 + g_2^2}{8} \rightarrow \frac{1}{8} \bigg( \frac{m_S^2 + |M_S|^2 + B_S }{m_S^2 + |M_S|^2 + B_S + 4|m_{D1}|^2} g_Y^2 + \frac{m_T^2 + |M_T|^2 + B_T }{m_T^2 + |M_T|^2 + B_T + 4|m_{D2}|^2} g_2^2 \bigg).
\end{align}
 This latter effect manifests itself as mixing terms in the Higgs mass matrix (\ref{eq:MassScalar}) $\Delta_{hs}, \Delta_{ht}$; provided that the other soft masses are large enough the suppression can be avoided. We may then ask how large these soft masses can be while preserving naturalness, to which we consider the ratiative corrections to $m_{H_{u,d}}^2$:
\begin{align}
\delta m_{H_{u,d}}^2 \supset& -\frac{1}{16\pi^2} ( 2 \lambda_S^2 m_S^2 + 6 \lambda_T^2 m_T^2) \log \left(\frac{\Lambda}{\mathrm{TeV}}\right)
\end{align}
where $\Lambda$ is the UV cutoff of the theory. Using $\Delta \equiv \frac{\delta m_h^2}{m_h^2 } $ then we have 
\begin{align}
m_S \lesssim& \mathrm{TeV} \left( \frac{1}{\lambda_S} \right) \left( \frac{\log \Lambda/\mathrm{TeV}}{3}\right)^{-1/2} \left(\frac{\Delta^{-1}}{20\%}\right)^{-1/2}   \nn\\
m_T \lesssim& 5\ \mathrm{TeV} \left( \frac{0.1}{\lambda_T} \right) \left( \frac{\log \Lambda/\mathrm{TeV}}{3}\right)^{-1/2} \left(\frac{\Delta^{-1}}{20\%}\right)^{-1/2}  .
\end{align}
The latter value is particularly important, because we also have the constraint that the triplet scalar expectation value must be small in order to avoid a large $\rho$ parameter. We have (see appendix)
\begin{align}
\Delta \rho \simeq& \frac{v^2}{(m_T^2 + |M_T|^2 + B_T + 4|m_{D2}|^2)^2} ( g_2 m_{D2} c_{2\beta} + \sqrt{2} \tilde{\mu} \lambda_T )^2 \lesssim 8 \times 10^{-4} 
\end{align}
 which is satisfied for \emph{any} value of $m_{D2}$ for $m_T \gtrsim 1.4\ \mathrm{TeV}$; but interestingly is also satisfied for any value of $m_T$ if $m_{D2} \gtrsim 1.4\  \mathrm{TeV}$. 

An interesting choice for the parameters $\lambda_S, \lambda_T$ would be for them to take their $N=2$ supersymmetric values at some scale, $\lambda_S = g_Y/\sqrt{2}, \lambda_T = g_2/\sqrt{2} $, although corrections due to the running would break this exact relation. However, even if the $N=2$ scale is intermediate (such as $10^{12} \gev$) assuming a desert we would still find $\lambda_S \simeq 0.2, \lambda_T \simeq 0.4$ at the low energy scale. In this case, in order to preserve naturalness we would require
\begin{align}
m_S \lesssim& 2 \mathrm{TeV} \left( \frac{0.2}{\lambda_S} \right) \left( \frac{\log \Lambda/\mathrm{TeV}}{20}\right)^{-1/2} \left(\frac{\Delta^{-1}}{20\%}\right)^{-1/2}   \nn\\
m_T \lesssim& 0.5\ \mathrm{TeV} \left( \frac{0.4}{\lambda_T} \right) \left( \frac{\log \Lambda/\mathrm{TeV}}{20}\right)^{-1/2} \left(\frac{\Delta^{-1}}{20\%}\right)^{-1/2}  .
\end{align}
In order to satisfy the tree-level $\Delta \rho$ constraint, we would either need to work at small $c_{2\beta}$, or take $m_{D2} \gtrsim 1.4\  \mathrm{TeV}$ and ensure that the ensuing Higgs mixing allows a large enough Higgs mass. We leave exploring this interesting possibility to future work: in this paper we shall take large values of $m_T$  and small values of $\lambda_T$ to keep the loop-level corrections to $\Delta \rho$ small. 

One final naturalness-related issue in these models is that the requirement of small $A$-terms typically diminishes stop mixing; in the case of large $\tan \beta$ the mixing is almost eliminated. Hence 
the contribution from the stops to $\Delta \rho$ is \cite{Djouadi:1996pa,Heinemeyer:2004gx,Barbieri:2006bg}:
\begin{align}
\Delta \rho^{\rm stops} \simeq& \frac{3\alpha_{em}}{16 \pi M_W^2 s_W^2} F_0 (m_{\tilde{t}_L}^2, m_{\tilde{b}_L}^2) \nn\\
\simeq&   \frac{\alpha_{em}}{16 \pi M_W^2 s_W^2} \frac{m_t^4}{m_{\tilde{t}_1}^2} \simeq 4\times 10^{-4} \left( \frac{500 \gev}{m_{\tilde{t}_1}} \right)^2
\end{align}
(where $F_0 (a,b)\equiv a+b - \frac{2ab}{a-b} \log \frac{a}{b}$) which is similar to the experimental value. 
In ``natural SUSY'' MSSM models the stops are lighter than about $600 \gev$, and so the stops by themselves will not be able to lift the Higgs mass to $125 \gev$, but have a large impact on the electroweak precision corrections. On the other hand, in the context of $\lambda$SUSY \cite{Barbieri:2006bg} the model can remain natural for stops as heavy as $1.5$ TeV \cite{Barbieri:2006bg,Hall:2011aa} because the relative contribution of stops to the Higgs mass compared to the tree-level effect is small. Clearly $\lambda_S$ in our models is the same as the $\lambda$ in $\lambda$SUSY.
Hence the simplest natural scenario is to take small $\tan \beta$, small $\lambda_T$, and $\lambda_S \sim 1$ with $m_T \sim $ few TeV and $m_S \lesssim $ TeV. The Dirac gaugino masses $m_{D3}, m_{D2}$ can be naturally large, but the Dirac Bino mass will be bounded above by the amount of mixing that it induces between the singlet and lightest Higgs (thus reducing the Higgs mass) to be a few hundred GeV. As a result of this, an amusing feature is that natural Dirac gauginos will lead to a \emph{Majorana} neutralino, since there will be non-negligible mixing between the Bino and the neutral Higgsinos. 

It is therefore worthwhile giving an example of such a ``natural'' model. In the context of ``dynamical $\mu$'' models we take $\lambda_S = -1.0, B\mu =8.0 \times 10^4 \gevsq, v_S = 170 \gev$ and $m_{D1} = -170 \gev, m_{D2} = m_{D3} = 10^3 \gev, m_O = 3$ TeV, $t_S = -1.5 \times 10^7 (\mathrm{GeV})^3, B_S = -5 \times 10^5 \gevsq, v_T = 0.465 \gev, m_T^2 = 2.2 \times 10^7 \gevsq$ while the first two generations have soft mass squareds of $4 \times 10^7 \gevsq$, with third generation soft masses at $(500 \gev)^2$,
which leads to light neutral Higgs masses (at one loop) of $121, 362, 429, 518 \gev$, a light pseudoscalar Higgs mass of $438$, a light 
charged Higgs mass of $412 \gev $, neutralinos of masses $82, 134,217,267,1056,1056 \gev$ and charginos of masses 
$121\gev $ and a TeV. We then find $\Delta \rho = 8 \times 10^{-4},\mathrm{BR}(b \rightarrow s\gamma) = 3.2 \times 10^{-4}$. We shall now turn to a discussion of the Higgs signatures at the LHC of natural Dirac gaugino models.

\section{Higgs production and branching ratios}
\label{SEC:DECAYS}

It is now important to consider the production cross-sections and branching ratios of the Higgs. 
In our Dirac gaugino models there is a singlet scalar which may mix with the lightest Higgs state, 
so we shall consider the branching ratios into each channel taking into account the mixing, 
providing some approximate expressions and then comparing them to the output of the \SPheno code created by \SARAH. 

We shall use the standard definitions
\begin{align}
R_{i} \equiv&  \frac{ BR( h \rightarrow ii)  }{BR_{SM}( h \rightarrow ii) } = \bigg| \frac{A_{ii}}{A^{SM}_{ii}} \bigg|^2\nn \nn\\
\mu_{ii} \equiv& \frac{\sigma (pp \rightarrow h) }{\sigma_{SM} (pp \rightarrow h) } R_{i}
\end{align}
where $A_{ii}$ is the amplitude.

To take Higgs mixing into account, consider the rotation of the states $h, H, s_R$ via a matrix $S$ so that 
\begin{align}
\thv[h, H, s_R] =& S . \thv[h_1, h_2, h_3].
\end{align}
We shall assume throughout that the lightest Higgs field $h_1$ has mass $125 \gev$, i.e. there is no additional lighter singlet. 
In this notation, we can then calculate how the production and decay channels are modified. At $125$ GeV the production cross-sections (as listed on the CERN yellow pages \cite{CernYellow}) are
\begin{align}
\sigma_{SM} (pp \rightarrow h) =& 19.5^{+14.7\%}_{-14.7\%} {\rm pb}\qquad \mathrm{gluon\ fusion} \nn\\
+&1.578^{+2.8\%}_{-3\%} {\rm pb}\qquad \mathrm{vector\ boson\ fusion}  \nn\\
+&0.6966^{+3.7\%}_{-4.1\%} {\rm pb}\qquad \mathrm{WH\ process} \nn\\
+&0.3943^{+5.0\%}_{-5.1\%} {\rm pb}\qquad \mathrm{ZH\ process} \nn\\
+&0.1302^{+11.6\%}_{-17.1 \%} {\rm pb}\qquad \mathrm{ttH\ process} \nn
\end{align} 
Of the initial eigenstates, only $h$ couples at tree level to the vector bosons. The coupling to gluons 
and in the $ttH$ process is proportional to the top quark coupling squared, so 
\begin{equation}
\Gamma( gg \rightarrow h_1) \sim  h_1 t\bar{t} \propto | S_{11} + S_{21} \cot \beta|^2.
\end{equation}
 Hence we can write
\begin{align}
F_g \equiv \frac{\Gamma(gg \rightarrow h_1)}{\Gamma_{SM}(gg \rightarrow h_1)} \simeq& \frac{ (19.5 +0.1302)| S_{11} + S_{21} \cot \beta|^2 + (1.578 + 0.6966 + 0.3943) | S_{11}|^2}{22.2991} \nn\\
=& 0.88 | S_{11} + S_{21} \cot \beta|^2 + 0.12 | S_{11}|^2 .
\end{align}
We can use the same approach for the decay branching ratios:
\begin{align}
R_{g}, R_c \propto   h t\bar{t} \sim& | S_{11} + S_{21} \cot \beta|^2 \nn\\
R_{b} \propto \Gamma (h_1 \rightarrow \bar{b} b) \sim& | S_{11} - S_{21} \tan \beta|^2 \nn\\
R_{\tau} \propto \Gamma (h_1 \rightarrow \bar{\tau} \tau) \sim& | S_{11} - S_{21} \tan \beta|^2 \nn\\
R_{W} \propto \Gamma (h_1 \rightarrow W W^*) \propto& |S_{11}|^2 \nn\\
R_{Z} \propto \Gamma (h_1 \rightarrow ZZ^*) \propto& |S_{11}|^2.
\label{EQ:roughRs}\end{align}
Since the photon couples at one loop to the Higgs, and the singlet couples to charged fields,
 the expression for the coupling to the photon will be more complicated. 
At $m_h = 125$ GeV, the standard model Higgs branching ratio (as listed on the CERN yellow pages \cite{CernYellow}) is dominated by
\begin{align}
BR_{SM}(h_1 \rightarrow \bar{b} b) =& 5.77 \times 10^{-1} \nn\\
BR_{SM}(h_1 \rightarrow W W^*) =& 2.15 \times 10^{-1} \nn\\
BR_{SM}(h_1 \rightarrow gg) =& 8.6 \times 10^{-2} \nn\\
BR_{SM}(h_1 \rightarrow \tau\tau) =& 6.3 \times 10^{-2} \nn\\
BR_{SM}(h_1 \rightarrow \bar{c} c) =& 2.91\times 10^{-2} \nn\\
BR_{SM}(h_1 \rightarrow ZZ^*) =& 2.6 \times 10^{-2}
\end{align}
So we can write
\begin{align}
\mu_{XX} =& F_g \frac{R_X (1 - \Br ( h \rightarrow \rm {invisible}))}{\sum_Y R_Y \Br_{SM} (h \rightarrow Y)}\nn\\
\simeq& F_g \frac{  R_X (1 - \Br ( h \rightarrow \rm {invisible}))}{ 0.640| S_{11} - S_{21} \tan \beta|^2 + 0.241 |S_{11}|^2 + 0.115 | S_{11} + S_{21} \cot \beta|^2} .
\label{EQ:MasterMu}\end{align}

To approximate the diphoton channel, we require the expression 
\begin{align}
R_{\gamma} =& \bigg| \frac{1}{A^{SM}_{\gamma\gamma}} \frac{v}{2}\bigg [ \frac{g_{h_1 VV}}{m_V^2} Q_V^2 A_1 (\tau_V)   + \frac{2 g_{h_1 ff} N_C Q^2_f }{m_f} A_{1/2} (\tau_f)+ \frac{g_{h_1 SS} N_C Q^2_S }{m_S^2} A_0 (\tau_S)  \bigg] \bigg|^2
\end{align}
where the functions $A_s$ where $s$ is the spin of the field in the loop are standard 
and given in \cite{Spira:1995rr}; $\tau_X \equiv 4 m_X^2/m_{h_1}^2$. For fermions and scalars they 
are well approximated by the large mass limits of $4/3$ and $1/3$ respectively, but 
for the W boson and top quark the values are $A_1 (\tau_W) \simeq -8.32, (4/3) A_{1/2} (\tau_t) \simeq 1.84$. If 
we consider that the singlet couples to some charged Dirac fermion we can write for the couplings $g_{h_1 ii}$
\begin{align}
g_{h_1 VV} =& \frac{2M_V^2}{v} S_{11} \nn\\
g_{h_1 t \bar{t}} =& \frac{m_t}{v} (S_{11} + \cot \beta S_{21}) \nn\\
g_{h_1 b \bar{b}} =& \frac{m_b}{v} (S_{11} - \tan \beta S_{21})
\end{align}
We can then consider enhancements through various fields remaining light, taking the mixing into account. 

\subsection{Charginos}
\label{sec:charginos}
The chargino mass matrix is expanded by new charged states from the triplet:
 in the basis $(T^+,\tilde{W}^+, H_u^+) / (T^-,\tilde{W}^-, H_d^+ )$ it is
\begin{equation}
\label{eq:CharginoMatrix}
M_{Ch} = 
\left(\begin{array}{c c c}
M_T + \frac{v_S \lambda_{ST}}{\sqrt{2}}    & m_{2D} + g_2v_T &\lambda_T  v c_\beta  \\
m_{2D} - g_2v_T & M_2   &  g_2 v s_\beta /\sqrt{2}\\
- \lambda_T  v s_\beta  & g_2 v c_\beta/\sqrt{2} & \tilde{\mu} - \sqrt{2}\lambda_T v_T \\
\end{array}\right) 
\end{equation}
 With $v_T \simeq 0$ and defining $\tilde{M}_T \equiv M_T + \frac{v_S \lambda_{ST}}{\sqrt{2}}$ this has determinant
\begin{align}
\det (M_{Ch}) = \frac{1}{4} \bigg[ -4 (m_{2D}^2 - M_2 \tilde{M}_T) \tilde{\mu} + 2\sqrt{2} g_2 \lambda_T m_{2D} v^2 c_{2\beta} + ( \lambda_T^2 M_2 - 2 g_2^2 \tilde{M}_T) v^2 s_{2\beta} \bigg].
\end{align}
Since the loop function $A_{1/2}$ varies very little between the lower bound on chargino masses 
from LEP ($105$ GeV, or $92$ GeV with caviats) and infinite mass, it is very well approximated 
by $4/3$ over all cases of interest. Then we can well approximate
\begin{align}
A_{\gamma\gamma}^{\rm Charginos} = &  \sum_f S_{11} \frac{v g_{hff}}{m_f} A_{1/2} (\tau_f) +S_{21} \frac{v g_{Hff}}{m_f} A_{1/2} (\tau_f) +S_{31} \frac{v g_{s_Rff}}{m_f} A_{1/2} (\tau_f) \nn\\
\simeq&  \frac{4}{3} v \bigg[ S_{11} \partial_h \log \det m_f + S_{21}  \partial_H \log \det m_f + S_{31}  \partial_{s_R} \log \det m_f \bigg] \nn\\
\simeq  \frac{4}{3}\frac{v}{\det M_{Ch}}  \bigg[ \frac{v}{2} S_{11} &\bigg(2\sqrt{2} g_2 \lambda_T m_{D2} c_{2\beta} + (2M_2 \lambda_T^2 - g_2^2 \tilde{M}_T^2) s_{2\beta} \bigg) \nn\\
  - \frac{v}{2} S_{21} &\bigg(2\sqrt{2} g_2 \lambda_T m_{D2} s_{2\beta} - (2M_2 \lambda_T^2 - g_2^2 \tilde{M}_T^2) s_{2\beta} \bigg)\nn\\
 - \frac{1}{\sqrt{2}} S_{31} &\bigg( \lambda_S(m_{D2}^2 - M_2 \tilde{M}_T)  + \lambda_{ST} (\frac{1}{4} g_2^2 v^2 s_{2\beta}- M_2 \tilde{\mu})\bigg) \bigg]
\end{align}
In the limit $M_2 = M_T = \lambda_{ST} = 0$, this simplifies to
\begin{align}
A_{\gamma\gamma}^{\rm Charginos} \simeq& \frac{4}{3} \frac{1}{\sqrt{2}m_{D2} \tilde{\mu} - g_2v^2 \lambda_Tc_{2\beta}} \bigg[ 2 g_2 \lambda_T v^2 (-c_{2\beta} S_{11} + s_{2\beta} S_{21}) + \lambda_S v m_{D2} S_{31} \bigg].
\label{EQ:APPcharginos}\end{align}

\begin{figure}
\begin{center}
\includegraphics[width=0.5\textwidth]{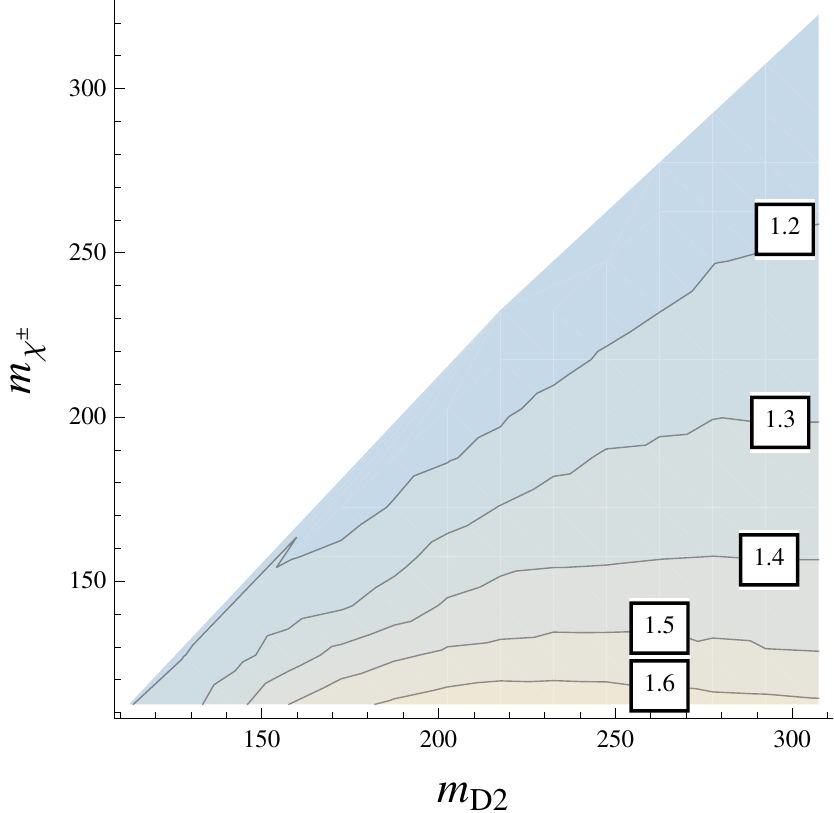}\includegraphics[width=0.5\textwidth]{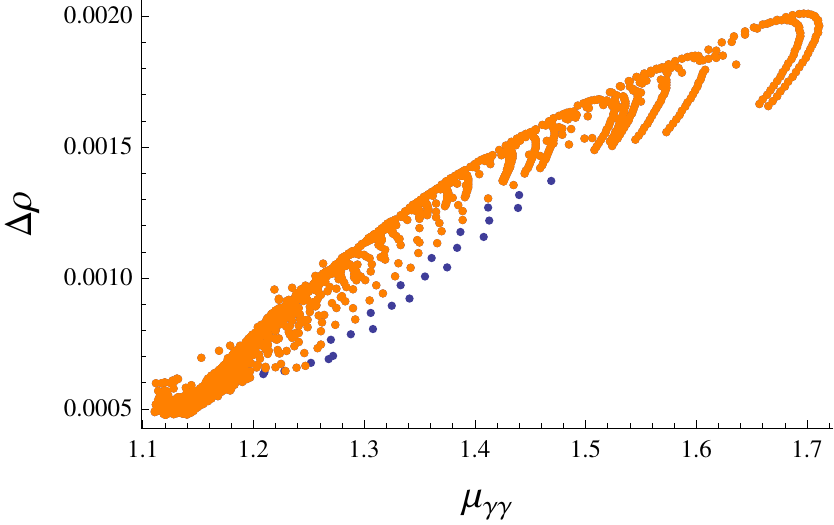}
\caption{Results of a scan over a subspace of the ``MSSM in disguise'' with  $\lambda_T = 0.7$, $\tan \beta = 50$, varying $m_{D2}$ and $\mu$ in order to vary the lightest chargino mass which is restricted to be Dirac Wino-like (above the contour the lightest chargino is Higgsino-like). The stop and sbottom masses were taken to be equal and varied in order to keep the Higgs mass at $125 \pm 4$ GeV (within reasonable accuracy of the results of the \SARAH-produced \SPheno code, which cannot include two-loop effects due to the Dirac gluino). Other non-zero soft parameters were: $m_{D1}= 200\ \rm{GeV}, m_{D3} = 1000\ \rm{GeV}, \lambda_S = 0.1$; the slepton and first two generations of squark masses squared were $4 \times 10^7 \gevsq$; the singlet and triplet scalar masses were approximately $5000\gev$; $B\mu$ was varied over $[300,5\times 10^5 ] \gevsq$. Left: contours of $\mu_{\gamma \gamma}$ (GeV). Right: $\mu_{\gamma \gamma}$ vs $\Delta \rho$; orange points have the lightest chargino mass greater than $105$ GeV, 
blue ones smaller. }\label{MSSMdisguisedgamma}
\end{center}
\end{figure}

The scenario of light charginos is particularly appropriate for the MSSM in disguise; since
 the adjoint scalars are all massive the main phenomenological difference with the MSSM will 
be the charginos and neutralinos. Hence this represents one useful limit of this formula, 
where there is negligible mixing between the Higgs states (i.e. $S_{11} \simeq 1, S_{i1} \simeq 0\ \forall\ i \ne 1$).
 Then it is clear that for an appreciably large coupling $\lambda_T$ and large $c_{2\beta}$ the $\gamma \gamma$
 channel can be enhanced \emph{without affecting the other channels}. We performed at scan at $\lambda_T = 1, \tan \beta = 50$
 varying $\mu$, and plot the contours of the Higgs to diphoton branching ratio in figure \ref{MSSMdisguisedgamma}, 
showing also the effect on $\Delta \rho$. This analysis is similar in spirit to models adding a triplet to the 
MSSM  \cite{Delgado:2012sm}, except instead of including a Majorana Wino and supersymmetric triplet mass we have included a Dirac Wino mass.

\begin{figure}[!h]
\begin{center}
\includegraphics[width=0.5\textwidth]{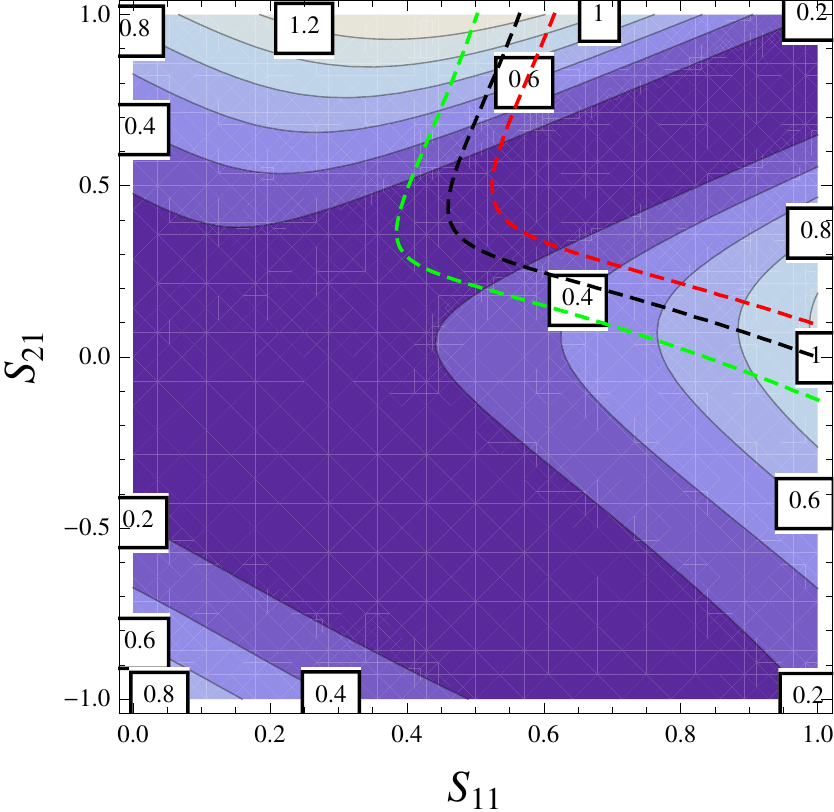}
\caption{Contours of $\mu_{bb} $ or equivalently $\mu_{\tau\tau}$ with $\tan \beta = 1.2$. The dashed lines represent $\mu_{WW} = 1.3$ (red) , $1$ (black) and $0.7$ (green).}
\label{FIG:btaucontour}\end{center}
\end{figure}

\begin{center}
\begin{figure}[!h]
\includegraphics[width=0.5\textwidth]{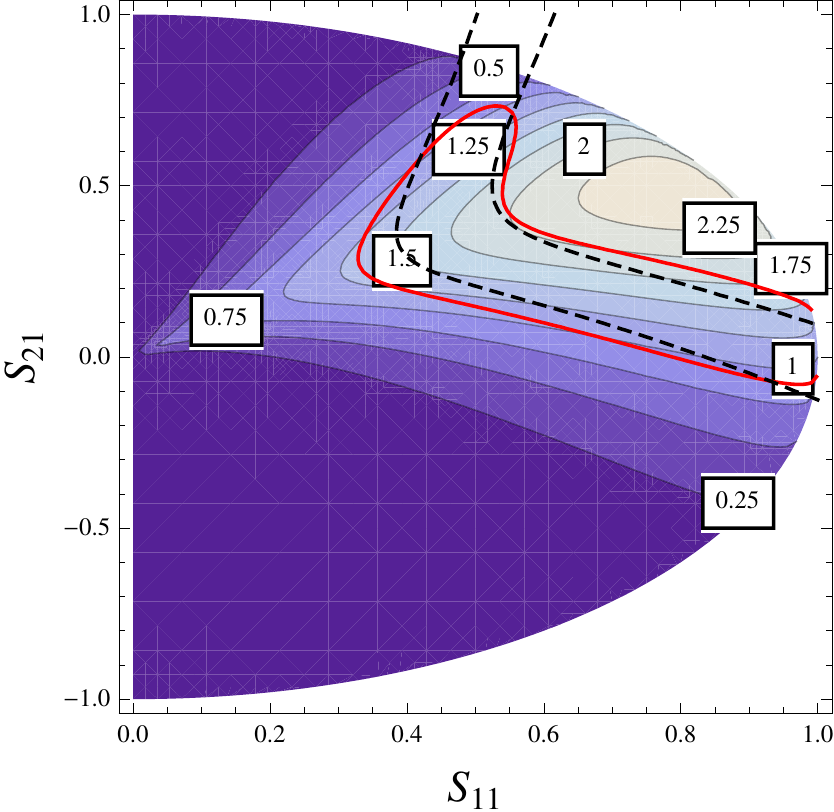}\includegraphics[width=0.5\textwidth]{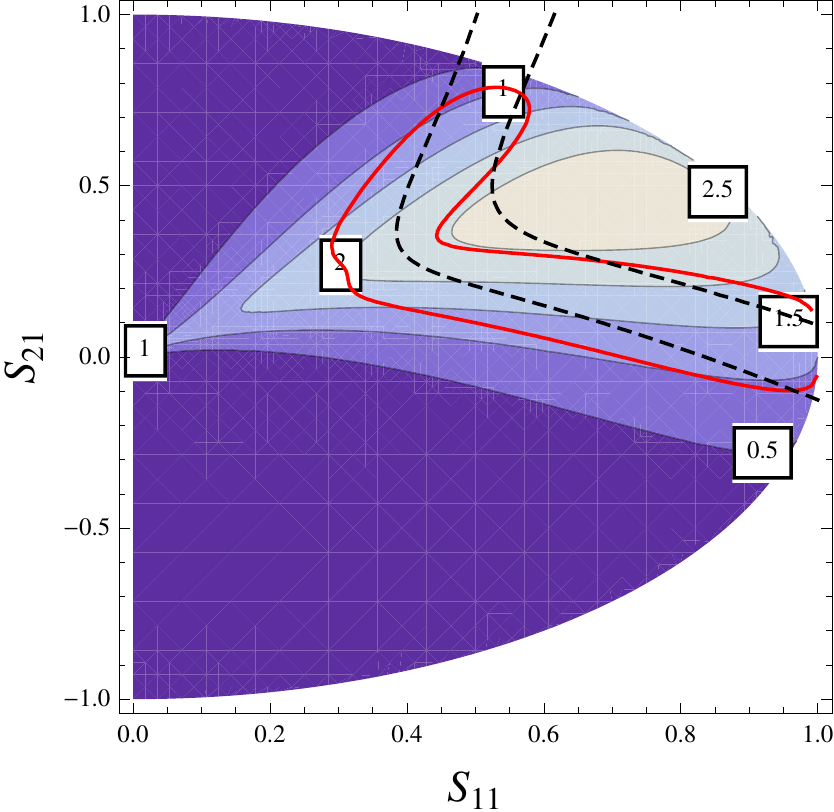}
\caption{Contours of $\mu_{\gamma \gamma}$ for $\lambda_S v/m_{\chi^+} = -2 $ (left) and $-3$ (right) with $\tan \beta = 1.2$.
 The black dashed lines are contours of $\mu_{WW} = 1.0 \pm 0.3$, and the red contour is a rough $95 \%$ confidence-level
 preferred region according to the data given in appendix \ref{APP:DATA} (excluding electroweak precision data, which is
 dependent on the spectra of any other light particles).}
\label{FIG:mixcharginolambdas}
\end{figure}
\end{center}

\begin{center}
\begin{figure}[!h]
\includegraphics[width=0.5\textwidth]{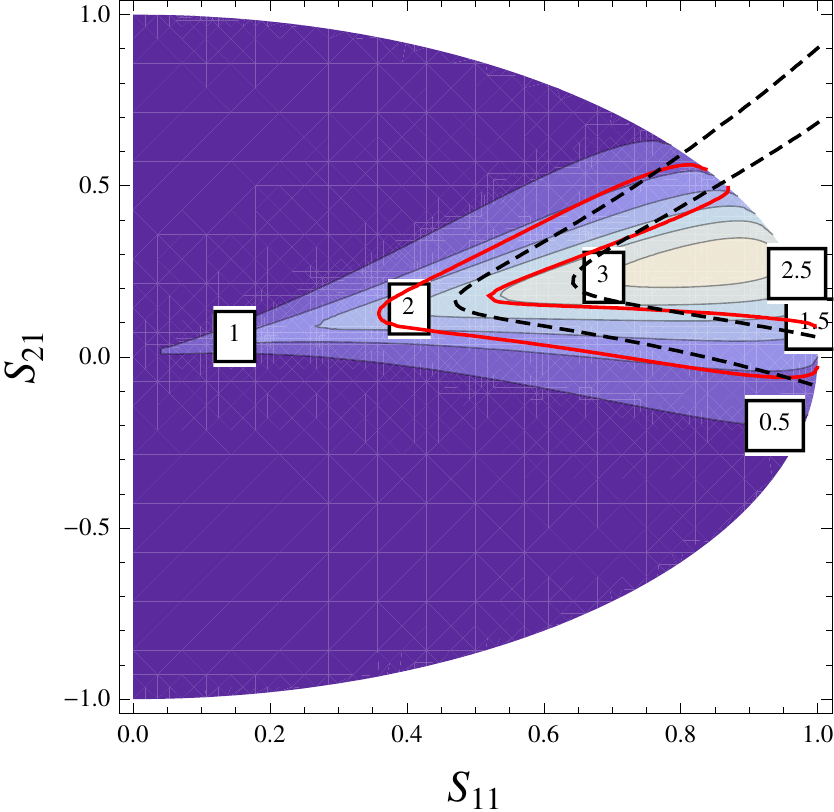}\includegraphics[width=0.5\textwidth]{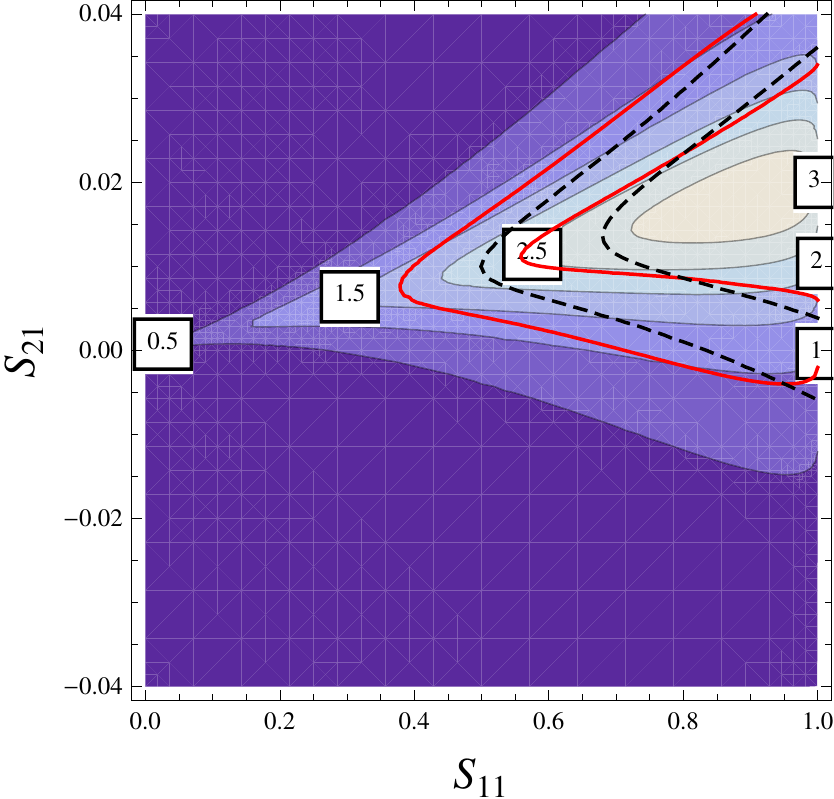}
\caption{Contours of $\mu_{\gamma \gamma}$ for $\lambda_S v/m_{\chi^+} = -3 $, $\tan \beta = 3$ (left) and $50 $ (right).
 The black dashed lines are contours of $\mu_{WW} = 1.0 \pm 0.3$, and the red contour is a rough $95 \%$ confidence-level 
preferred region according to the data given in appendix \ref{APP:DATA} (excluding electroweak precision data, which is 
dependent on the spectra of any other light particles).}
\label{FIG:mixcharginotans}
\end{figure}
\end{center}

An alternative application of formula (\ref{EQ:APPcharginos}) is to allow appreciable Higgs mixing but
take the large $m_{D2}$ limit, leaving light Higgsinos. We then find
\begin{align}
\label{eq:higgsinos}
R_\gamma \simeq& \bigg| S_{11} - 0.28  \cot \beta S_{21} - 0.15  \frac{ v\lambda_S}{\mu} S_{31} \bigg|^2.
\end{align}
Hence the diphoton production rate can be enhanced for suitably large $v \lambda_S/\mu$ and $S_{31}$. 
This is applicable for the NMSSM too, and is particularly interesting because by varying the Higgs 
mixing terms we can simultaneously enhance $\mu_{\gamma \gamma}$, decrease the $\mu_{b\bar{b}}$
 and $\mu_{\tau\bar{\tau}}$ while maintaining $\mu_{WW}$ and $\mu_{ZZ}$ roughly unchanged if so desired. 
 This is similar to singlet extensions of the MSSM without Dirac gauginos \cite{SchmidtHoberg:2012yy} and 
can be easily understood from equations (\ref{EQ:roughRs}) and (\ref{EQ:MasterMu}): 
by having a small positive admixture of $H$ we enhance the coupling of the Higgs to tops, and hence to gluons; by reducing
$S_{11}$ we reduce the coupling to bottoms and taus, which reduces the total width of the Higgs - both of these can compensate for the
reduction in coupling to $W$s. In figure \ref{FIG:btaucontour} we show how $\mu_{bb}$ and $\mu_{\tau\tau}$ are affected by the mixing. The above effect of the mixing and charginos on $\mu_{\gamma \gamma}$ is then illustrated in figures \ref{FIG:mixcharginolambdas} and \ref{FIG:mixcharginotans}.

\subsection{Charged Higgs}

The charged Higgs fields can also contribute to the diphoton rate in these models. The full charged Higgs matrix 
involves not just the $H^\pm$ of the MSSM, but also the charged triplet scalars; however, from the $\rho$-parameter 
constraint we know that the triplet scalars must be heavy (see section \ref{SEC:NATURAL}), so we shall neglect their contribution. Hence we can approximate
\begin{align}
A_{\gamma\gamma}^{\rm Charged\ Higgs} \simeq \frac{v^2}{3 m_{H^{\pm}}^2} \frac{1}{16} \bigg[& - S_{11} \bigg(g_Y^2 -3 g_2^2 + 2 \lambda_S^2 - 14 \lambda_T^2 + (g_Y^2 + g_2^2 - 2(\lambda_S^2 + \lambda_T^2)) \cos 4 \beta \bigg) \nn\\
 &+ S_{21} (g_Y^2 + g_2^2 - 2(\lambda_S^2 + \lambda_T^2)) \sin 4 \beta  \nn\\
&+ \frac{8 S_{31}}{v} \bigg( g_Y m_{D1} c_{2\beta} + \sqrt{2} \lambda_S ( \tilde{\mu} + \frac{1}{\sqrt{2}} \kappa v_S s_{2\beta} + \frac{1}{2} A_S s_{2\beta}\bigg)\bigg]
\end{align}
A large contribution to the diphoton rate from charged Higgs loops can then arise when mixing between the lightest Higgs and the singlet is substantial. Note that light charged Higgs fields in the limit of very 
small $\tan\beta$ also often demand light stops and charginos to cancel large contributions to $b\to s\gamma$.

\subsection{Stops and staus}

Of the squarks in the theory, the stops and staus - having the strongest couplings to the light Higgs - have been 
studied in the MSSM as candidates to modify the Higgs to diphoton rate \cite{Carena:2012gp,Sato:2012bf}. In models with Dirac gauginos, the new D-term 
contributions to the potential modify the squark masses; for the stop and stau their mass matrices are given 
in appendix equations (\ref{FullSquarkMassesStop}) and (\ref{FullSquarkMassesSbottom}).
Neglecting the normal D-term components, $v_T$ and the $A$-terms gives approximately 
\begin{align}
 \C{M}_{\tilde{t}}^2 \simeq& \2b2[m_Q^2 + \frac{1}{3} g_Y m_{D1} v_S + v^2 \frac{1}{2} y_t^2  s_\beta^2, -\frac{1}{\sqrt{2}} v y_t   \tilde{\mu} c_\beta][-\frac{1}{\sqrt{2}} v y_t   \tilde{\mu} c_\beta, m_{t_R}^2 - \frac{4}{3} g_Y m_{D1} v_S + v^2 \frac{1}{2} y_t^2 s_\beta^2 ] \nn\\
\C{M}_{\tilde{\tau}}^2 =& \2b2[m_L^2  - g_Y m_{D1} v_S+ v^2 \frac{1}{2} y_\tau^2 c_\beta^2 , -\frac{1}{\sqrt{2}} v y_\tau \tilde{\mu} s_\beta][-\frac{1}{\sqrt{2}} v y_\tau \tilde{\mu} s_\beta, m_{\tau_R}^2 +2g_Y m_{D1} v_S+ v^2  \frac{1}{2} y_t^2c_\beta^2 ].
\end{align}
It is straightforward to derive expressions for the the couplings of the stops and staus to the Higgs eigenstates, but since the full expressions are lengthy we give here simplified formulae neglecting subleading terms proportional to $M_Z$ and setting the $A$-terms and $v_T$ to zero:
\begin{align}
A_{\gamma\gamma}^{\rm Stops} \simeq \frac{4 m_t^2}{9 m_{\tilde{t}_1}^2 m_{\tilde{t}_2}^2} \bigg[ &S_{11} \Big(m_{\tilde{t}_1}^2 + m_{\tilde{t}_2}^2 - \tilde{\mu}^{2} \cot^{2}\beta    \Big) + S_{21}\Big(m_{\tilde{t}_1}^2 + m_{\tilde{t}_2}^2 + \tilde{\mu}^{2} \Big)\cot\beta \\
&- S_{31}\bigg(\frac{1}{\sqrt{2}}  v \lambda_S \tilde{\mu} \cot^{2}\beta +  \frac{m_{D1}M_Z s_W}{3m_t^2}  \Big(3 m_{t}^{2}  +4 m_Q^2  - m_U^2  \bigg) \bigg)\bigg] \nn\\
A_{\gamma\gamma}^{\rm Staus} \simeq  \frac{m_\tau^2}{3 m_{\tilde{\tau}_1}^2 m_{\tilde{\tau}_2}^2} \bigg[ & S_{11} \Big(m_{\tilde{\tau}_1}^2 + m_{\tilde{\tau}_2}^2 - \tilde{\mu}^{2} \tan^{2}\beta  \Big) - S_{21}  \Big(m_{\tilde{\tau}_1}^2 + m_{\tilde{\tau}_2}^2 + \tilde{\mu}^{2} \Big)\tan\beta  \\
&-  S_{31} \bigg( \frac{1}{\sqrt{2}}  v \lambda_S \tilde{\mu} \tan^{2}\beta  -\frac{m_{D1}M_Z s_W}{m_\tau^2}  \Big( 2 m_L^2  - m_E^2  + m_{\tau}^{2} \bigg) \bigg) \bigg].\nn
\end{align}
Here $m_{\tilde{t}_i},m_{\tilde{\tau}_i}  $ are the masses of the stop and stau eigenstates respectively. The Dirac mass enters into the above by shifting the mases (e.g. $m_{\tilde{t}_1}^2 + m_{\tilde{t}_2}^2 = m_Q^2 + m_U^2 + 2 m_t^2 - g_Y m_{D1} v_S$) but clearly only plays a significant role in enhancing the diphoton channel if there is large mixing of the lightest Higgs with the singlet scalar.

\section{MSSM without $\mu$-term}
\label{SEC:MULESS}

\begin{figure}[hbt]
\begin{minipage}{0.95\linewidth}
\includegraphics[width=0.45\linewidth]{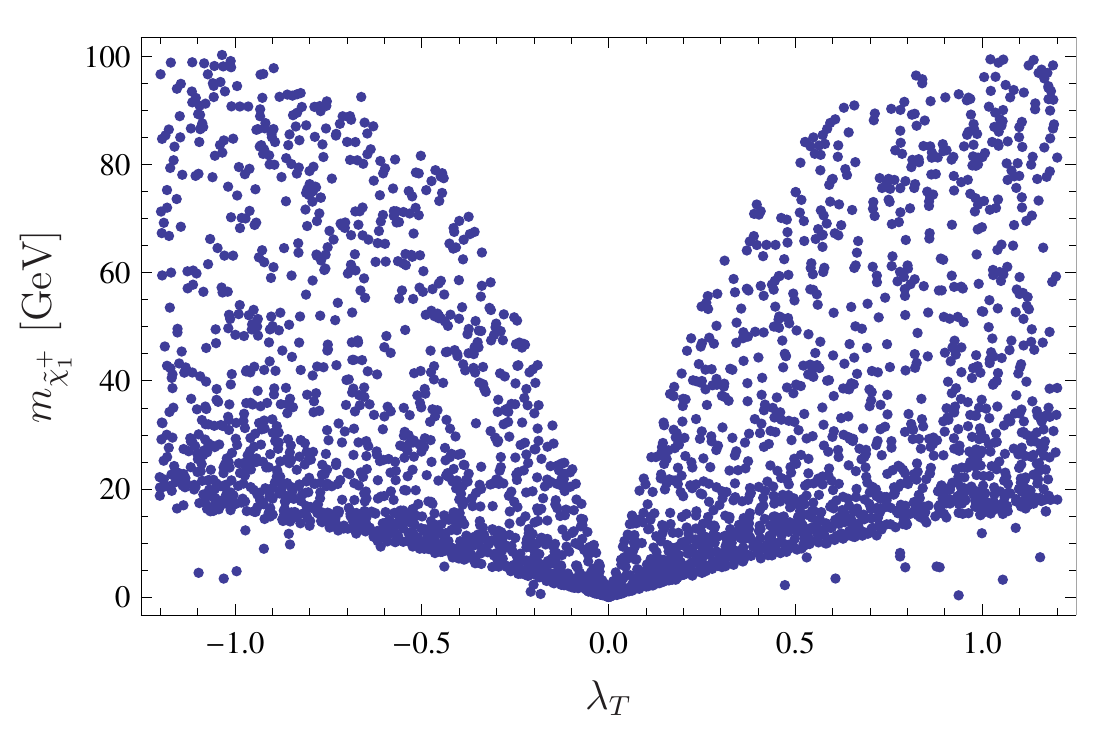} \hfill
\includegraphics[width=0.45\linewidth]{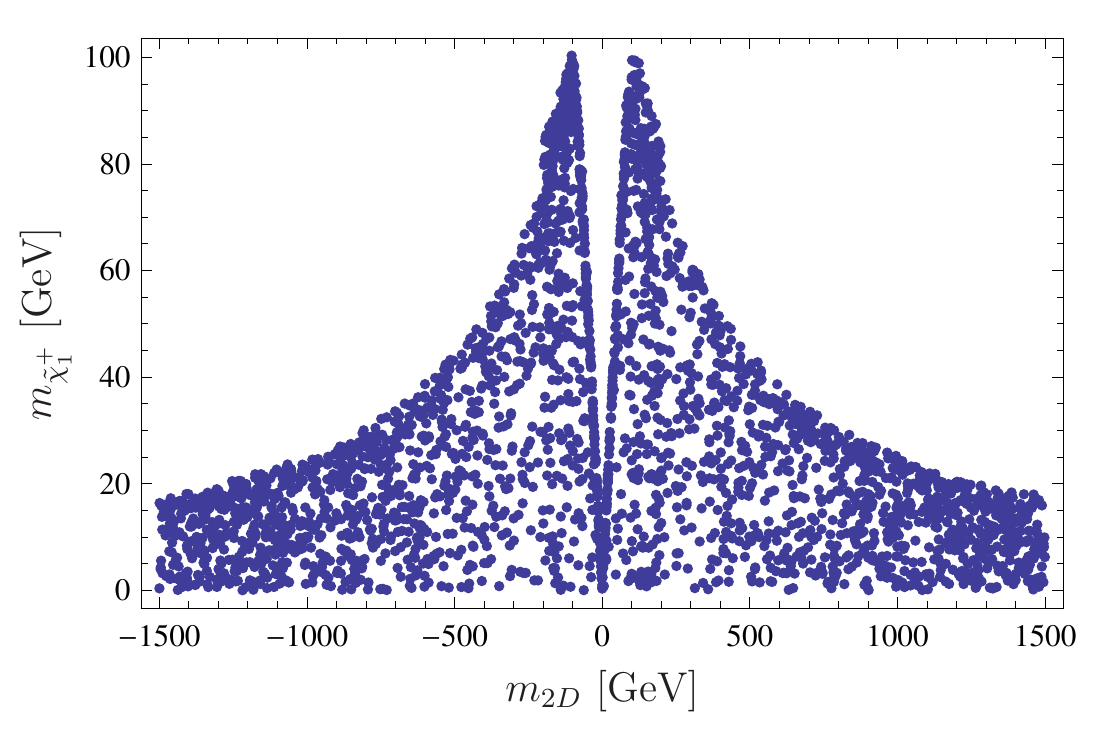} 
\end{minipage}
\caption{Mass of the lightest charginos as function of $\lambda_T$ (left) and $m_{2D}$ (right). The others parameters 
have been chosen in the ranges: 
$\tan\beta=[30,70]$, $\lambda_S = [-0.2,0.2]$, $\lambda_{ST} = \kappa = [-1.7,1.7]$, $B_S = B_T = [-10^7,10^7]$~GeV, 
$B_\mu = [-100,100]$~GeV, $v_T = v_S = [-1,1]$~GeV, $m_{1D} = [-1.5,1.5]$~TeV, $T_{top} = [-1.5,1.5]$~TeV. 
The sfermion sector is fixed by $m_{q,ii}^2 = m_{d,ii}^2 = m_{u,ii}^2 = 5\cdot 10^6~\text{GeV}^2$, 
$m_{e,ii}^2 = m_{l,ii}^2 = 5\cdot 10^5~\text{GeV}^2$ ($i=1,2$), $m_{q,33}^2 = m_{d,33}^2 = m_{u,33}^2 = 10^6~\text{GeV}^2$, 
$m_{e,33}^2 = m_{l,33}^2 = 10^5~\text{GeV}^2$.}
\label{fig:noMu_C1mass}
\end{figure}

In the presence of Dirac gaugino masses it is possible to remove the Higgsinos' mass term from 
the superpotential of the MSSM \cite{Nelson:2002ca}. This is another way of curing the intrinsic $\mu$-problem of the 
MSSM. Furthermore, an approximate $U(1)_R$ symmetry naturally guarantees that $\tan\beta$ is large, 
explaining the top/bottom quark mass hierarchy. In contrast to its appealing theoretical aspact, 
the $\slashed{\mu}$SSM is under substantial pressure from experimental data. First, LEP put a lower limit on 
the mass of the lightest chargino of 94~GeV \cite{Nakamura:2010zzi}. The chargino mass eq.~(\ref{eq:CharginoMatrix})
reads in this limit
 \begin{equation}
M_{Ch} = 
\left(\begin{array}{c c c}
0  & m_{2D} &\lambda_T  v c_\beta  \\
m_{2D}  & 0   &  g_2 v s_\beta /\sqrt{2}\\
- \lambda_T  v s_\beta  & g_2 v c_\beta/\sqrt{2} & 0\\
\end{array}\right) 
\end{equation}
It is known that it is possible to fulfill this bound by a careful choice of $\lambda_T$ and $m_{2D}$: 
a large value of $\lambda_T$ as well as $m_{2D}$ around 107~GeV is needed to maximize the mass of the 
lightest chargino, see Fig.~\ref{fig:noMu_C1mass}. The highest mass which can be reached at tree-level 
is about 110~GeV. This is also not 
improved at the one-loop level because the loop corrections due to the heavy triplets even tend to 
decrease the mass. Hence, the highest mass we could find in our scans calculating the full one-loop spectrum was 103~GeV. 

\begin{figure}[hbt]
\begin{minipage}{0.95\linewidth}
\includegraphics[width=0.45\linewidth]{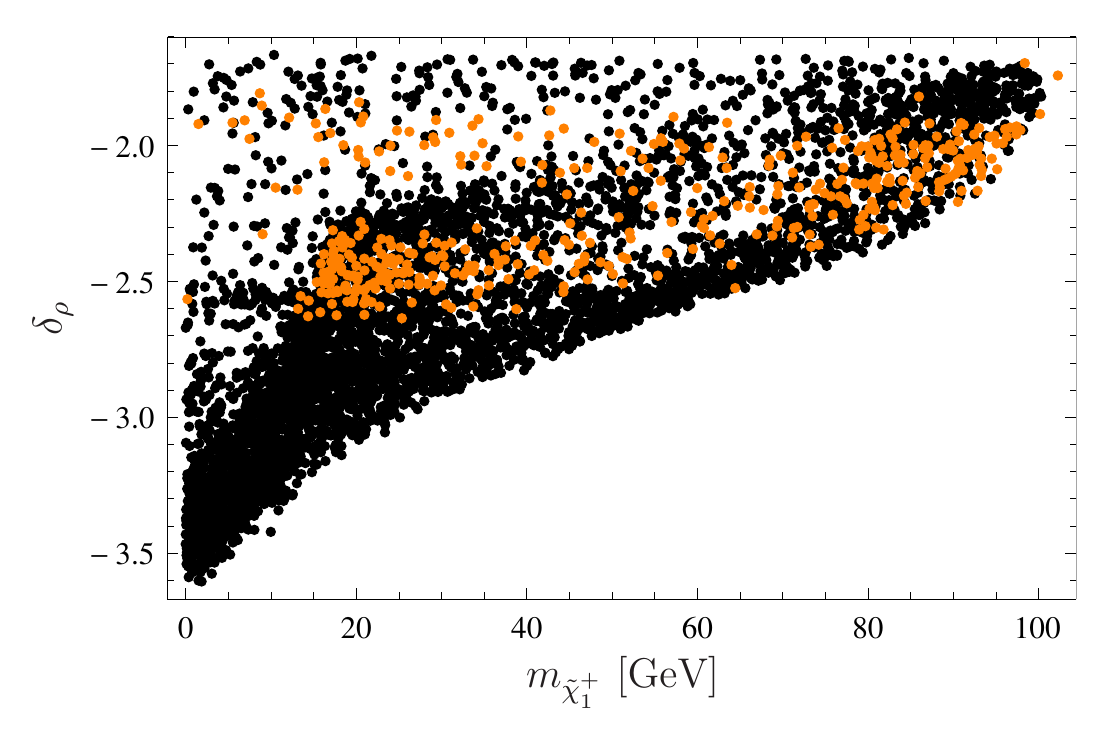}  \hfill
\includegraphics[width=0.45\linewidth]{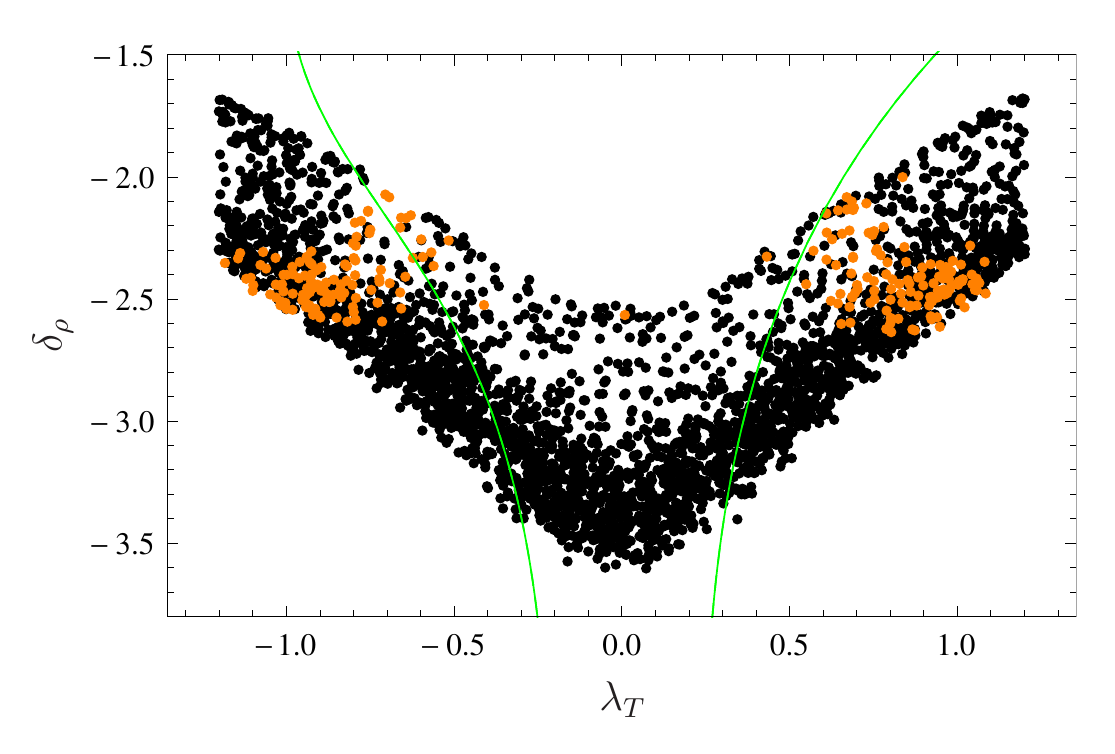}  
\end{minipage}
\caption{Left: $\delta_\rho\equiv \log_{10} \Delta \rho$ versus the lightest chargino mass. The orange points indicate a mass of the 
lightest Higgs in the range of 122--128~GeV. Right: $\delta_\rho$ as function of $\lambda_T$. The green 
line shows the result using the approximative formula given in ref.~\cite{Nelson:2002ca}. The input values 
are the same as for Fig.~\ref{fig:noMu_C1mass}. }
\label{fig:noMu_deltaRho}
\end{figure}

A more severe problem is that the large values of $\lambda_T$ lead often to a huge contribution to $\Delta \rho$. 
The reason is that this coupling breaks the custodial $SU(2)_L$ symmetry present in the SM and MSSM: especially,
 the chargino and neutralino-loops contribute differently to the $W$ and $Z$ self-energies. The large impact was
 already pointed out in ref.~\cite{Nelson:2002ca} using an approximation for the resulting contributions to 
the $T$-parameter. We repeat this analysis with the full numerical evaluation of $\Delta \rho$. We find a 
strong correlation in this model between the mass of the lightest chargino and the smallest possible value 
of $\Delta \rho$ as depicted in Fig.~\ref{fig:noMu_deltaRho}. 

Even if the full calculations leads to somewhat smaller values of $\Delta \rho$ than the approximate one 
for $\lambda_T > 1$, all points which fulfill the limit of $m_{\tilde{\chi}^+_1}>94$~GeV suffer from a 
large $\Delta \rho$ of at least 0.003. Many points are even above 0.01. Note that $\Delta \rho$ is 
very quickly increasing with the chargino mass. Therefore, demanding $\Delta \rho < 0.0008$ would rule out 
all points with chargino masses above 20~GeV. Furthermore, it can also be seen that in general the points 
with a Higgs mass between 122 and 128~GeV lead independently from the chargino mass to $\Delta \rho > 0.001$. 
The reason is that a sizable contribution of $\lambda_T$ to the tree-level Higgs mass is needed. This could, 
of course, be circumvented to some extent by allowing for even larger values of $T_{top}$. However, this is 
in contradiction to the approximate $U(1)_R$-symmetry which suppresses in general the trilinear terms. 
Therefore, if we restrict ourself to moderate values of the squared 
squark mass parameters and trilinear soft-term, this model is always in conflict with $\Delta \rho$: 
even if it would be somehow possible to find kinematical configurations to significantly reduce the LEP 
limits on the chargino mass, the now existing bound on the Higgs mass still predicts too large values of $\Delta \rho$.  

\section{Dynamical $\mu$ models}
\label{SEC:RBREAKING}

Although the ``MSSM without $\mu$ term'' may be severely challenged, by allowing a substantial expectation 
value for the singlet the various problems can be cured. The Higgs potential will always lead to a 
non-zero value for $v_S$ as can be seen from the minimisation condition (\ref{eq:minVS}), but in order for 
this to be significant we can allow a non-zero negative value for $B_S$ and/or a non-zero tadpole 
term $t_S$. Both of these are generically present  and do not break R-symmetry. In this 
scenario, as in the MSSM without $\mu$ term, we take the only significant source of R-symmetry breaking to be a $B\mu$ term. 

This scenario is particularly interesting from the perspective of Higgs mixing, since the 
singlet adjoint scalar will typically be light - we see from the minimisation conditions 
that the parameter $\tilde{m}_S^2 $ in the tree-level mass-squared matrix (\ref{eq:MassScalar}) is
\begin{align}
\tilde{m}_S^2 = - \frac{\sqrt{2} t_S + v_0^3}{v_S}
\end{align}
and, expecting from naturalness and RGE running \cite{Goodsell:2012fm} $t_S \sim v_0^3 \sim v^3, v_S \sim v$, $\tilde{m}_S^2 \sim v^2 $. 
There may thus be substantial mixing between the singlet and original ``$h$'' eigenstate; the 
size of $B\mu$ term then controls the amount of ``$H$'' in the lightest Higgs mass eigenstate. 

Moreover, the singlet couples to the gauginos via the coupling $\lambda_S$, which, if $m_{D2}$ 
is not small, will be predominantly Higgsino-like. This then offers the possibility of realising 
the scenario considered in section \ref{sec:charginos}. We have therefore conducted a scan over a portion of 
the parameter space of these models, concentrating on models with a small component of mixing 
between $h$ and $H$ but substantial $S_{11}$ and $S_{31}$ copmonents, using the \SPheno code 
produced by \SARAH. $\tan \beta$ was taken to be $1.5$ and $\lambda_S$ was varied from a negative initial value in order to 
fix the Higgs mass at $125 \pm 4 \gev$ - recall that this is a rather conservative error range.  The other parameters 
varied were $m_{D1} \in [-800,800]\gev, v_S \in [130,430]\gev, B\mu \in [312,90312]\gevsq, $ while 
the non-zero fixed soft parameters were 
$\lambda_T = 0.021, B_S=-5\times 10^5 \gevsq, t_S = -1.5\times 10^7 \gev^3, m_T^2 = 2.5\times 10^7 \gevsq,
 m_O^2 = 9 \times 10^6 \gevsq, m_{D2} = 600 \gev, m_{D3} = 1000 \gev$. The slepton and first two generations of 
squark masses squared were $4 \times 10^7 \gevsq$ while the third generation squark masses squared were $1.5 \times 10^6 \gevsq $.

\begin{figure}
\begin{center}
\begin{tabular}{cc}
\includegraphics[width=0.45\textwidth]{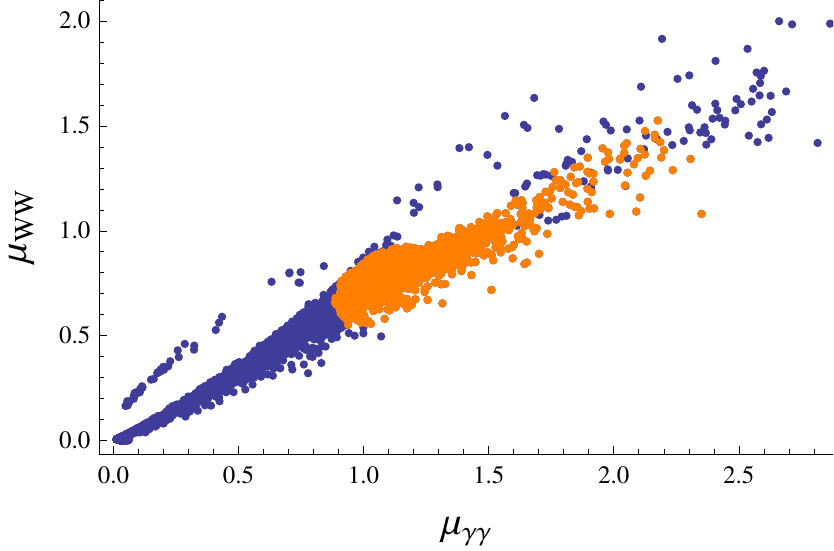} & \includegraphics[width=0.45\textwidth]{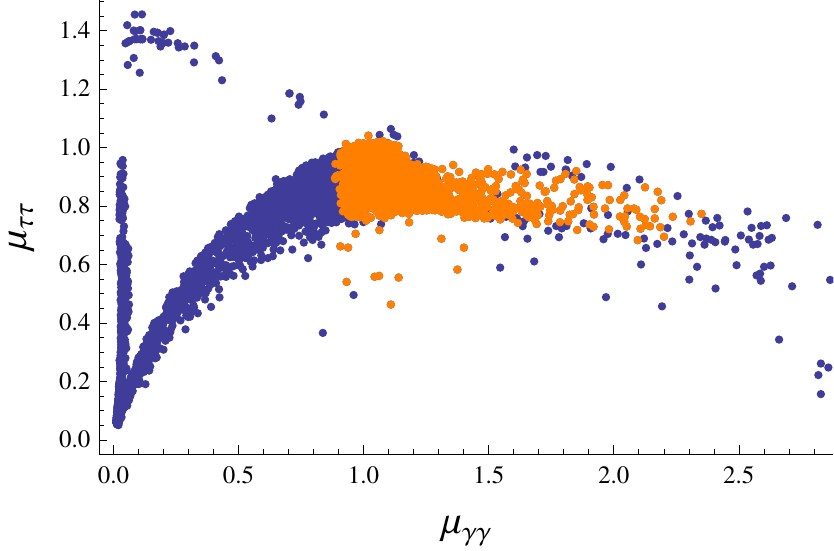} \\
\includegraphics[width=0.45\textwidth]{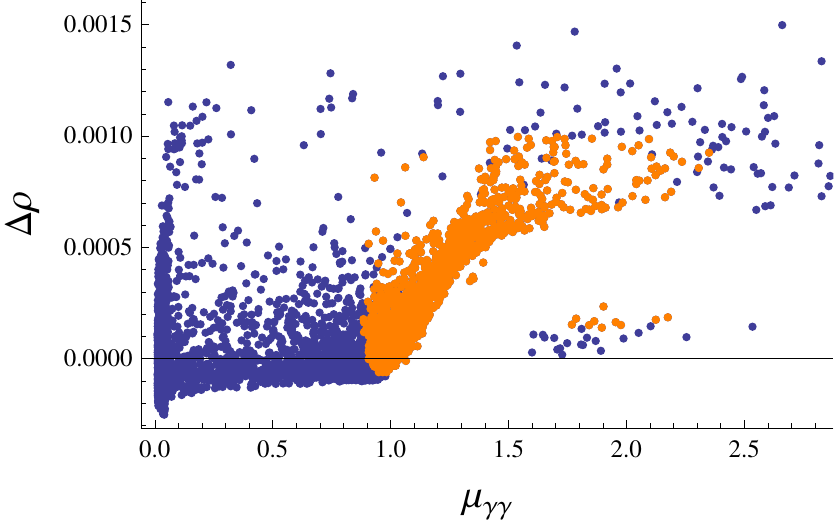} & \includegraphics[width=0.45\textwidth]{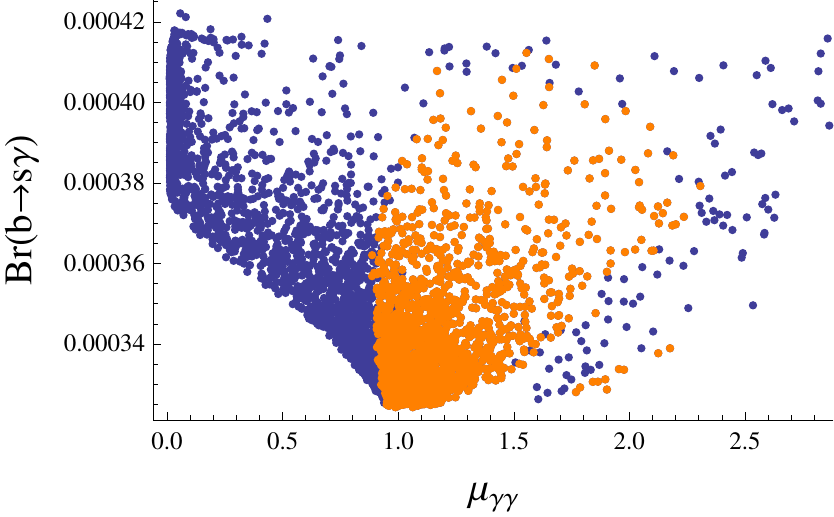}
\end{tabular}
\caption{Production cross-section times branching ratio for a scan of a subspace of the ``dynamical $\mu$'' scenario, 
with parameters chosen to enhance Higgs mixing. In the scan, $\tan \beta$ was taken to be $1.5$ and $\lambda_S$ was varied in 
order to fix the Higgs mass at $125 \pm 4 \gev$ with inital value being negative. The other parameters varied 
were $m_{D1} \in [-800,800]\gev, v_S \in [130,430]\gev, B\mu \in [312,90312]\gevsq, $ while the non-zero fixed soft 
parameters were $\lambda_T = 0.021, B_S=-5\times 10^5 \gevsq, t_S = -1.5\times 10^7 \gev^3, 
m_T^2 = 2.5\times 10^7 \gevsq, m_O^2 = 9 \times 10^6 \gevsq, m_{D2} = 600 \gev, m_{D3} = 1000 \gev$.
 The slepton and first two generations of squark masses squared were $4 \times 10^7 \gevsq$ while the third
 generation squark masses squared were $1.5 \times 10^6 \gevsq $. The points shown in orange pass all experimental 
limits and furthermore lie within a crude $95\%$ confidence limit via $\chi^2$ based on the current Higgs and electroweak 
precision data as described in appendix \ref{APP:DATA}.}
\end{center}
\label{HiddenRScans}\end{figure}

\begin{center}
\begin{figure}
\includegraphics[width=0.5\textwidth]{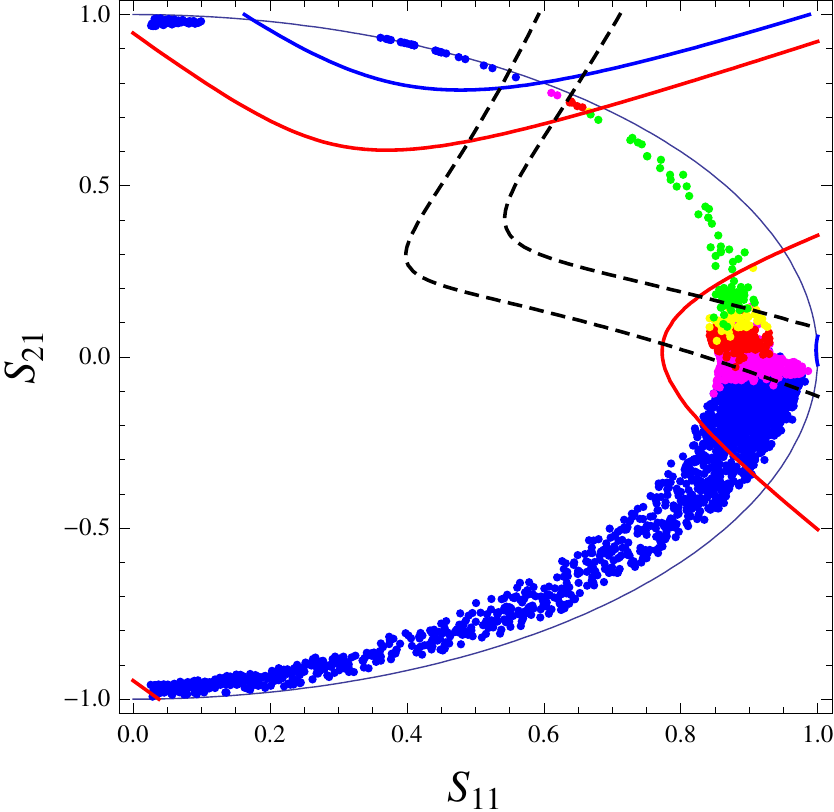}\includegraphics[width=0.5\textwidth]{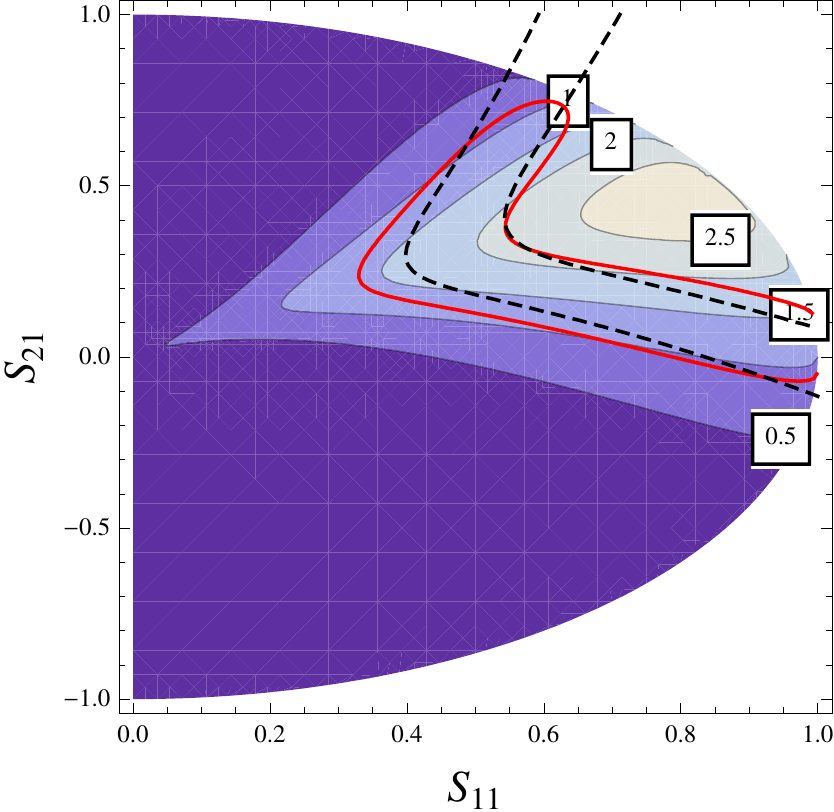}
\caption{Left: subset of results of the ``dynamical $\mu$'' scan shown in figure \ref{HiddenRScans} plotted in terms of the 
mixing matrix parameters $S_{11}$ and $S_{21}$ where all point pass experimental limits (except Higgs branching ratios). 
The point colours denote the $\mu_{\gamma \gamma}$ values, varying from dark blue for $\mu_{\gamma \gamma} < 1 $ to red 
with $1.33 < \mu_{\gamma \gamma} < 1.67$ to green for $\mu_{\gamma \gamma} > 2 $. The solid contour lines show 
$\mu_{bb} = 0.6, 1.0$ (thick red, thick blue respectively) and $S_{11}^2 + S_{21}^2 = 1$ (thin blue) while the 
dashed black contours show $\mu_{WW} = 1.0 \pm 0.3$. 
Right: shown for comparison, this is a plot of the same form as figure \ref{FIG:mixcharginolambdas} but with $\tan \beta = 1.5, \lambda v/m_f = -2$.}
\label{HiddenRmixing}\end{figure}
\end{center}

Figure \ref{HiddenRScans}
shows the $\mu_{WW}$ and $\mu_{\tau \tau}$ 
values versus $\mu_{\gamma \gamma}$ ($\mu_{ZZ}$ and $\mu_{bb}$ being, as per the approximate formulae, 
almost identical to $\mu_{WW}$ and $\mu_{\tau \tau}$ respectively) while also giving the values 
of $\Delta \rho$ and $\mathrm{BR}(b \rightarrow s\gamma)$. As can be seen from the plots, there are many 
experimentally viable model points. To further elucidate the comparison with our predicted scenario, a 
plot of the mixing parameters $S_{11}$ and $S_{21}$ with $\mu_{\gamma \gamma}$ revealed by the colour 
of the points is shown in figure \ref{HiddenRmixing}. 

It is worthwhile to pick out one example point; this has $\lambda_S = -1.3, B\mu =4.18 \times 10^4 \gevsq, v_S = 162 \gev$ and $m_{D1} = -152 \gev$ 
which leads to light neutral Higgs masses of $122, 227, 394 \gev$, a light pseudoscalar Higgs mass of $285$, a light 
charged Higgs mass of $210 \gev $, neutralinos of masses $65, 156,220,315,649,651 \gev$ and charginos of masses 
$144,647,652 \gev $. The mixing data is  $S_{11} = 0.89, S_{21} = 0.08, S_{31}=-0.44$ which results 
in $\mu_{\gamma \gamma}=1.6, \mu_{WW} = 1.14, \mu_{ZZ} = 0.97, \mu_{bb}=0.81, \mu_{\tau\tau} =0.86  $ 
and $\Delta \rho = 7 \times 10^{-4},\mathrm{BR}(b \rightarrow s\gamma) = 3.6 \times 10^{-4}$.

\subsection*{Models with light stops}

In the previous scan, we held the third generation masses to be heavy to be above search bounds and to diminish their contribution to $\Delta \rho$ whilst still remaining natural. However, it is also straightforward to find models of the above class that have light stops, which would be natural even for the MSSM but also interesting for LHC searches. Taking the same fixed values as above except now with third generation soft masses at $(500 \gev)^2, m_{D2} = 1 \mathrm{TeV},\tan \beta =2$,  one example has $\lambda_S = -0.96, B\mu =3.5 \times 10^4 \gevsq, v_S = 193 \gev$ and $m_{D1} = -294 \gev$ 
which leads to $v_T = 0.46 \gev$, light neutral Higgs masses of $122, 256, 360 \gev$, a light pseudoscalar Higgs mass of $290$, a light 
charged Higgs mass of $254 \gev $, neutralinos of masses $115,138,319,343\gev$ and charginos of masses 
$129,1055,1059 \gev $. The mixing data is  $S_{11} = 0.97, S_{21} = 0.03, S_{31}=-0.23$ which results 
in $\mu_{\gamma \gamma}=1.4, \mu_{WW} = 1.2, \mu_{ZZ} = 1.0, \mu_{bb}=1.0, \mu_{\tau\tau} =1.08  $ 
and $\Delta \rho = 4.8 \times 10^{-4},\mathrm{BR}(b \rightarrow s\gamma) = 3.4 \times 10^{-4}$. 

Another example with more mixing has $\lambda_S = -1.14, B\mu =4.6 \times 10^4 \gevsq, v_S = 178 \gev$ and $m_{D1} = -283 \gev$ 
which leads to $v_T = 0.47 \gev$, light neutral Higgs masses of $121, 290, 380 \gev$, a light pseudoscalar Higgs mass of $330$, a light 
charged Higgs mass of $283 \gev $, light neutralinos of masses $115,151,318,354 \gev$ and charginos of masses 
$140 \gev $ and a TeV. The mixing data is  $S_{11} = 0.95, S_{21} = 0.07, S_{31}=-0.30$ which results 
in $\mu_{\gamma \gamma}=1.6, \mu_{WW} = 1.3, \mu_{ZZ} = 1.1, \mu_{bb}=0.95, \mu_{\tau\tau} =1.0  $ 
and $\Delta \rho = 7\times 10^{-4},\mathrm{BR}(b \rightarrow s\gamma) = 3.4 \times 10^{-4}$.

A final example has $\tan \beta = 1.5, t_S = 1.2 \times 10^7 \gevsq, \lambda_S=-1.44, B\mu =4.7 \times 10^4 \gevsq, v_S = 144 \gev$ and $m_{D1} = -247 \gev$ 
which leads to $v_T = 0.3 \gev$, light neutral Higgs masses of $125.3, 218,389\gev$, a light pseudoscalar Higgs mass of $298\gev$, a light 
charged Higgs mass of $206 \gev $, light neutralinos of masses $89,153,301,368 \gev$ and charginos of masses 
$141 \gev $ and a TeV. The mixing data is  $S_{11} = 0.88, S_{21} = 0.09, S_{31}=-0.47$ which results 
in $\mu_{\gamma \gamma}=1.8, \mu_{WW} = 1.2, \mu_{ZZ} = 1.0, \mu_{bb}=0.80, \mu_{\tau\tau} =0.86  $ 
and $\Delta \rho = 8\times 10^{-4},\mathrm{BR}(b \rightarrow s\gamma) = 3.5 \times 10^{-4}$.

\section{Conclusions}
\label{SEC:CONCLUSIONS}

Dirac gaugino models are gaining increased interest as non-minimal supersymmetric standard models with enhanced naturalness compared to the MSSM and an enhanced Higgs mass that can also relax bounds on direct superpartner searches. With the latest update to the \SARAH package, it is now possible to study such models quantitatively using modern numerical tools, and this work is a first step in exploring phenomenologically the low-energy parameter space. We have discussed the properties of three different Dirac gaugino scenarios that are subclasses of the minimal Dirac gaugino extension of the (N)MSSM: the ``MSSM in disguise,'' the ``MSSM without $\mu$ term'' of \cite{Nelson:2002ca} and a new scenario involving a dynamical $\mu$ term. While the first of these is phenomenologically very similar to the MSSM with higher-dimensional operators, we found that the second is unfortunately severely challenged by the current data. The third scenario, on the other hand, can be particularly natural and also has many characteristics appropriate to allow Higgs mixing and thus modifications of the Higgs production and decay rates; in particular, it is possible for example to enhance the diphoton signal, suppress the bottom and tau signals, while leaving the Z and W channels roughly the same as the Standard Model. We have performed a first examination of its parameter space but clearly it would be interesting to examine it further, particularly as new Higgs data becomes available.

There are now many interesting directions for future work. One will be to compare specific models directly with collider data, particularly in the context of models with light stops. In addition, it would be interesting to see how embedding the models we have discussed in particular high-energy completions affects the discussion of naturalness. Furthermore, constraints due to dark matter (assuming a thermal history of the universe or otherwise) can now also be applied. On the technical side, to further refine the precision of the Higgs mass, the leading two-loop corrections involving the Dirac gluinos should now be calculated. This work is therefore one step on the increasingly attractive path of bringing the phenomenology of Dirac gauginos closer to the level of understanding of the (N)MSSM.

\section*{Acknowledgements}

We thank Jong Soo Kim, Nicolas Bernal and Werner Porod for fruitful discussions.
MDG was supported by ERC advanced grant 226371. KB is supported in part by the European contract “UNILHC” PITN-GA-2009-237920.

\appendix

\section{Tree-level parameters of the model}
\label{APP:TREE}

In this section we summarise the tree-level parameters of the model; see also \cite{Benakli:2011kz}. Here we add the tadpole term for the singlet and expressions for the stop and stau mass matrices including the new Dirac gaugino D-term corrections. 

\subsection{Higgs potential}

It will be useful to introduce the following effective mass parameters:
\begin{eqnarray}
\label{defmueff}
\tilde{\mu} & = & \mu +   \frac{1}{\sqrt{2}} (\lambda_S \,   v_S + \lambda_T  \,  v_T)
\nonumber \\
\tilde{B}{\mu} &=&B\mu +   \frac{\lambda_S}{\sqrt{2}} (M_S  + A_S) v_S+
\frac{\lambda_T}{\sqrt{2}} (M_T  + A_T) v_T +   \frac{1}{2} \lambda_S \,   \kappa  \,   v_S^2
\end{eqnarray}

\subsubsection{Equations of motion for the CP-even neutral fields}

The scalar potential for the CP-even neutral fields is  given by:
\begin{eqnarray}
V_{EW} &= &\left[ \frac{g^2+g'^2}{4}c_{2\beta}^2  \,   \,   + \,   \,  \frac{
\lambda_S^2 +  \lambda_T^2}{2}s_{2\beta}^2\right] \frac{v^4}{8} \nonumber \\ 
&&+\left[  m_{H_u}^2 s_{\beta}^2+ m_{H_d}^2 c_{\beta}^2 \,   \,  +\tilde{\mu}^2
- \tilde{B}{\mu}  \,   s_{2\beta}   \,   \,  + ( g  \,    m_{2D}    \,  v_T -
g'    m_{1D}   \,   v_S   )c_{2\beta} \right] \frac{v^2}{2} \nonumber \\ 
&&+
\sqrt{2} t_S v_S + \frac{\kappa^2}{4}v_S^4+\frac{\kappa}{\sqrt{2}}\frac{(3M_S+A_\kappa)}{3}  \,   v_S^3+
\frac{1}{2} \tilde{m}_{SR}^2   \,   v_S^2 + \,   \,   \frac{1}{2}
\tilde{m}_{TR}^2   \,   v_T^2 
\label{VEW1}
\end{eqnarray}
where the effective masses  for the real parts of the $S$ and $T$ fields read:
\begin{eqnarray}
\tilde{m}^2_{SR}& = & M_S^2+m_S^2+4 m^2_{1D}+ B_S, \qquad  \, \tilde{m}^2_{TR}= 
M_T^2+m_T^2+4 m^2_{2D}+ B_T 
 \end{eqnarray}
There is no restriction on the sign of the different  mass parameters $m_S^2$ and $B_S$ at this stage.

The imaginary parts of the fields have been dropped as their vevs are vanishing
due to the assumed CP conservation \cite{Belanger:2009wf}. The coefficients  of 
the corresponding  quadratic terms:
\begin{eqnarray}
 \tilde{m}_{SI}^2 =  M_S^2+m_S^2- B_S, \qquad  \qquad  \tilde{m}_{TI}^2=
M_T^2+m_T^2-B_T 
 \end{eqnarray}
do not, in contrast to the CP-even partners, receive contributions from
$D$-terms proportional to the Dirac masses.

As is customarily done for the (N)MSSM, the minimization of the scalar potential allows
here also to express $\tilde{\mu}$ and $\tilde{B}{\mu} $ as a function of the other parameters:
\begin{eqnarray}
\tilde{\mu}^2+  \frac{M_{Z}^2}{2}  = \frac{ m_{H_d}^2 - t_\beta^2 \,  \, 
m_{H_u}^2}{t_\beta^2 - 1} +\left[ \frac{  t_\beta^2 +1}{t_\beta^2 -1} \right]
\left( g  \,    m_{2D}    \,  v_T - g'    m_{1D}   \,   v_S \right)
\label{EQ:hmin}\end{eqnarray}
and
\begin{eqnarray}
M_A^2&\equiv&  \frac{2 \tilde{B}{\mu} }{s_{2\beta}}  =  2\tilde{\mu}^2+
m_{H_u}^2 + m_{H_d}^2- \frac{ \lambda_S^2 +  \lambda_T^2}{2} v^2c_{\beta}^2
\label{MA1}
\end{eqnarray}
The new equations are
\begin{align}
0 =& \kappa^2 v_S^3+ \frac{\kappa}{\sqrt{2}}(A_\kappa +3 M_S) v_S^2  + (\tilde{m}_{SR}^2 + \lambda_S (\lambda_S - \kappa s_{2\beta}) \frac{v^2}{2} )v_S + \sqrt{2} t_S + v_0^3 \nn\\
0=& (2\tilde{m}^2_{TR} + \lambda_T^2 v^2) v_T + v^2 [ g m_{2D}  c_{2\beta} +{\sqrt{2}} \tilde{\mu}   \lambda_T - \frac{\lambda_T}{\sqrt{2}} ( M_T+    A_T)  s_{2\beta} ] 
\label{EQ:vsvt0}\end{align}
where
\begin{eqnarray}
v_0^3 & = & - \frac{v^2}{2} \left[ g'    m_{1D} c_{2\beta}    -  \lambda_S  
\left( \sqrt{2} \mu - \frac{(A_S +M_S)} {\sqrt{2}}s_{2\beta} +  \lambda_T
v_T  \right) \right].
\end{eqnarray}

We can use these to solve for the masses in terms of the vevs. However, since the vev of $T$ contributes to the $W$ boson mass, the electroweak precision data give important bounds 
on the parameters of the model. For instance, using 
$\Delta \rho =(4.2 \pm 2.7) \times 10^{-4} $~\cite{Amsler:2008zzb,Aaltonen:2012bp,Abazov:2012bv,Group:2012gb,Barger:2012hr},  
we require:
\begin{equation}
\Delta \rho  \simeq 4 \frac{v_T^2}{v^2} <  1 \times 10^{-3} \ (95 \%)
\end{equation}
which is satisfied for $v_T\lesssim 4$ GeV. For large triplet masses we have
\begin{eqnarray}
v_T & \simeq &  \frac{v^2}{2\tilde{m}^2_{TR}} \ \ \left[ - g   
m_{2D}  c_{2\beta} -{\sqrt{2}} \tilde{\mu}   \lambda_T + \frac{\lambda_T
}{\sqrt{2}} ( 
M_T+    A_T)  s_{2\beta}  \right]
\end{eqnarray}

\subsubsection{Masses of  the CP even neutral scalars}

Introducting the notation 
\begin{eqnarray}
\tilde{m}_{S}^2&=&\tilde{m}_{SR}^2
+\lambda_S^2 \,  \frac{v^2}{2} - \kappa  \,   \lambda_S \,  \frac{v^2}{2 }
s_{2\beta}+3{\kappa^2}v_S^2+
 \frac{\sqrt{2}}{3}\kappa  \,  v_S \,   (A_\kappa +3 M_S) \nn\\
\tilde{m}_T^2 &=& \tilde{m}_{TR}^2 +\lambda_T^2  \frac{v^2}{2}
 \end{eqnarray}
the mass matrix for the CP even scalars in the basis $\{h, H,
S_R,T^0_R\}$ is:
\begin{eqnarray}
\label{eq:MassScalar}
\left(\begin{array}{c c c c }
M_Z^2+\Delta_h s_{2\beta}^2 & \Delta_h s_{2\beta}  c_{2\beta}  & \Delta_{hs}    
& 
\Delta_{ht} \\
\Delta_h s_{2\beta}  c_{2\beta} & M_A^2-\Delta_h s_{2\beta}^2   & \Delta_{Hs}    
&\Delta_{Ht}    \\
 \Delta_{hs}     & \Delta_{Hs}      & \tilde{m}_S^2  &  \lambda_S \lambda_T
\frac{v^2}{2} \\
  \Delta_{ht}     &\Delta_{Ht}    &  \lambda_S \lambda_T \frac{v^2}{2} & 
\tilde{m}_T^2  \\
\end{array}\right) 
\end{eqnarray}
where we have defined:
\begin{eqnarray}
\Delta_h&=&\frac{v^2}{2}(\lambda_S^2+\lambda_T^2)-M_Z^2 
\end{eqnarray}
which vanishes when $\lambda_S$ and $\lambda_T$ take their $N=2$ values 
\cite{Antoniadis:2006uj}.
We denote non-diagonal elements describing the mixing of $S_R$ and $T^0_R$ states 
with the light Higgs $h$:
\begin{eqnarray}
 \Delta_{hs} =- 2 \frac {v_ S} {v}   \tilde{m}_{SR}^2- \sqrt{2} \kappa 
\frac{v_S^2}{v}(A_\kappa +3 M_S)- 2 \kappa^2 \frac{v_S^3}{v}, \qquad  
 \Delta_{ht} = -2 \frac {v_ T} {v}   \tilde{m}_{TR}^2
\end{eqnarray}
while
\begin{eqnarray}
 \Delta_{Hs} = g' m_{1D} v  s_{2\beta}  - \lambda_S \frac{v(A_s +M_s)}{\sqrt{2}}
c_{2\beta},\qquad
 \Delta_{Ht}  =  - g m_{2D} v  s_{2\beta}  - \lambda_T \frac{v(A_T +M_T)}{\sqrt{2}}
c_{2\beta} \nonumber \\
 \end{eqnarray}
stand for the corresponding mixing with heavier Higgs, $H$.

Let us work with $M_S = M_T = A_S = A_\kappa = A_T = 0$. Rewriting the $v_S$ equation we have
\begin{align}
0 =& \kappa^2 v_S^3+ \frac{\kappa}{\sqrt{2}}(A_\kappa +3 M_S) v_S^2  + \tilde{m}_{SR}^2 v_S + \sqrt{2} t_S \nn\\
&- \frac{v^2}{2} \left[ g'    m_{1D} c_{2\beta}    -  \lambda_S  \left( \sqrt{2} \tilde{\mu} - \kappa s_{2\beta}v_S - \frac{(A_S +M_S)} {\sqrt{2}}s_{2\beta} \right) \right] \nn\\
\rightarrow& \kappa^2 v_S^3 + \tilde{m}_{SR}^2 v_S + \sqrt{2} t_S - \frac{v^2}{2} \left[ g'    m_{1D} c_{2\beta}    -  \lambda_S  \left( \sqrt{2} \tilde{\mu} - \kappa s_{2\beta} v_S\right) \right] = 0 \label{eq:minVS}
\end{align}
Then using the minimisation conditions we can write
\begin{align}
\Delta_{hs} =& v  [v_S\lambda_S (\lambda_S - \kappa s_{2\beta}) - g' m_{1D} c_{2\beta} + \sqrt{2} \lambda_S \mu + \lambda_S \lambda_T v_T] \nn\\
=& v[ \sqrt{2} \lambda_S \tilde{\mu} - g' m_{1D} c_{2\beta} - \lambda_S v_S \kappa s_{2\beta}]
\label{EQ:HSRestriction}\end{align}
Clearly $\Delta_{hs} \lesssim M_Z^2$ is necessary to prevent a see-saw reduction in the lightest Higgs mass.
We have
\begin{align}
\tilde{m}_{SR}^2 =& - \frac{1}{v_S} \bigg[ \sqrt{2} t_S + \kappa^2 v_S^3- \frac{v^2}{2}  g'    m_{1D} c_{2\beta}    +  \frac{v^2}{2}\lambda_S  \left( \sqrt{2} \tilde{\mu} - \kappa s_{2\beta} v_S\right)  \bigg] \nn\\
=&  - \frac{1}{v_S} \bigg[ \sqrt{2} t_S + \kappa^2 v_S^3 + \frac{v}{2} \Delta_{hs} \bigg]
\end{align}


\subsection{Squark masses}

The squark masses are modified by the Dirac mass terms via the D-term contribution. We give here the expressions for the mass matrices for stops, sbottoms and staus:
\begin{align}
(\C{M}_{\tilde{t}}^2)_{11} =& m_Q^2 + m_t^2 + \frac{1}{3} g_Y m_{D1} v_S + g_2 m_{D2} v_T +  M_Z^2 ( \frac{1}{2} - \frac{2}{3} s_W^2) c_{2\beta} \nn\\
(\C{M}_{\tilde{t}}^2)_{21} =& m_t (A_t s_\beta - \tilde{\mu} c_\beta)\nn\\
(\C{M}_{\tilde{t}}^2)_{22} =&m_{t_R}^2 + m_t^2 - \frac{4}{3} g_Y m_{D1} v_S +  \frac{2}{3} M_Z^2 s_W^2 c_{2\beta}
\label{FullSquarkMassesStop}
\end{align}
The entries of the sbottom mass matrix read
\begin{align}
(\C{M}_{\tilde{b}}^2)_{11} =& m_Q^2 +m_b^2  + \frac{1}{3} g_Y m_{D1} v_S - g_2 m_{D2} v_T + M_Z^2 ( -\frac{1}{2} + \frac{1}{3} s_W^2) c_{2\beta} \nn\\
(\C{M}_{\tilde{b}}^2)_{21} =& m_b (A_b c_\beta - \tilde{\mu} s_\beta)\nn\\
(\C{M}_{\tilde{b}}^2)_{22} =& m_{D}^2  +m_b^2 +\frac{2}{3}g_Y m_{D1} v_S - \frac{1}{3} M_Z^2 s_W^2 c_{2\beta}
\label{FullSquarkMassesSbottom}
\end{align}
The entries of the stau mass matrix read
\begin{align}
(\C{M}_{\tilde{\tau}}^2)_{11} =& m_L^2 +m_\tau^2 - g_Y m_{D1} v_S - g_2 m_{D2} v_T + M_Z^2 ( -\frac{1}{2} + s_W^2) c_{2\beta} \nn\\
(\C{M}_{\tilde{\tau}}^2)_{21} =& m_\tau (A_\tau c_\beta - \tilde{\mu} s_\beta)\nn\\
(\C{M}_{\tilde{\tau}}^2)_{22} =& m_{\tau_R}^2+m_\tau^2 + 2 g_Y m_{D1} v_S  -  M_Z^2 s_W^2 c_{2\beta}
\label{FullSquarkMasses}
\end{align}

\section{One-loop effective potential}
\label{APP:EFFPOT}

The standard general form of the Higgs potential up to quartic order is 
\begin{eqnarray}
V_{eff} & = & (m_{H_u}^2+\mu^2) |H_u|^2 +
(m_{H_d}^2+\mu^2) |H_d|^2
         - [m_{12}^2 H_u\cdot H_d + h.c. ] \nonumber \\
   &   & + \frac{1}{2}\big[\frac{1}{4}(g^2+g'^2) + \lambda_1\big]
                       (|H_d|^2)^2
         +\frac{1}{2}\big[\frac{1}{4}(g^2+g'^2) + \lambda_2\big]
                       (|H_u|^2)^2 \nonumber \\
   &  &  +\big[\frac{1}{4}(g^2-g'^2) + \lambda_3\big]
                       |H_d|^2 |H_u|^2
         +\big[-\frac{1}{2}g^2 + \lambda_4\big]
(H_d\cdot H_u)(H_d^*\cdot H_u^*)\nonumber\\
   & &   +\big(\frac{\lambda_5}{2} ( H_d\cdot H_u)^2
              +\big[ \lambda_6 |H_d|^2 + \lambda_7 |H_u|^2\big]
                    ( H_d\cdot H_u) + h.c. \big),
\end{eqnarray}
In Dirac gaugino models where we integrate out the heavy singlet and triplet scalars, we find $\lambda_3$ and $\lambda_4$ have a tree level contribution, so we write $\lambda_3 = 2\lambda_T^2 + \lambda_3^\prime, \lambda_4 = \lambda_S^2 - \lambda_T^2 + \lambda_4^\prime$ where $\lambda_3^\prime, \lambda_4^\prime$ are the loop corrections to the potential. Then we find at one loop $\lambda_4^\prime =\lambda_5 = \lambda_6 = \lambda_7 = 0,$ with the remaining contributions from the singlet and triplet scalars given by 

\input{lambdas.tex}

Note that there are also important contributions from the stops and sbottoms, which at this level are identical to those in the MSSM (hence we separated them in equation (\ref{EQ:EffHiggsMass})) except with the soft masses shifted by the Dirac gaugino D-term contributions as in equations (\ref{FullSquarkMassesStop}) and (\ref{FullSquarkMassesSbottom}).

\section{Experimental data}
\label{APP:DATA}

In this appendix we give the experimental data, current at time of writing, used in the text for crude $95\%$ confidence level limits. The Higgs mass reported by CMS \cite{CMS:2012gu} and ATLAS \cite{Atlas:2012gk} is:
\begin{equation}
m_H = 125.3 \pm 0.4 \pm 0.5\ \mathrm{GeV\ (CMS)}, 126.0 \pm 0.4 \pm 0.4\ \mathrm{GeV\ (ATLAS)}. 
\end{equation}

In table \ref{APP:tableofmus} we give the latest reported values of $\mu_{ii}$. 

\begin{table}[!h]
\begin{center}
$
\begin{array}{|c|c|c|c|}\hline
 & \mathrm{CMS} & \mathrm{ATLAS} & \mathrm{Tevatron} \\ \hline
\mu_{\gamma\gamma} & 1.6 \pm 0.4 & 1.8 \pm 0.5 & 3.62^{+2.96}_{-2.54}\\
\mu_{ZZ} & 0.64 \pm 0.57  (7\ \mathrm{TeV}) & 1.7\pm 1.1  (7\ \mathrm{TeV})   & \\ 
& 0.79 \pm 0.56 ( 8\ \mathrm{TeV}) & 1.3 \pm 0.8 \ (8\ \mathrm{TeV})  & \\
\mu_{WW} & 0.38 \pm 0.56 \ (7\ \mathrm{TeV}) & 0.5\pm 0.6 \ (7\ \mathrm{TeV})   & 0.32^{+1.13}_{-0.32} \\ 
& 0.98 \pm 0.71 \ (8\ \mathrm{TeV}) & 1.9 \pm 0.7 \ (8\ \mathrm{TeV})  & \\
 \mu_{bb} & 0.59 \pm 1.17 \ (7\ \mathrm{TeV} & 0.46\pm 2.18 \ (7\ \mathrm{TeV})   & 1.97^{+0.74}_{-0.68} \\ 
& 0.41 \pm 0.94 \ (8\ \mathrm{TeV}) & & \\
\mu_{\tau\tau} & 0.62 \pm 1.17 \ (7\ \mathrm{TeV}) & 0.45\pm 1.8 \ (7\ \mathrm{TeV})   & \\ 
& -0.72 \pm 0.97 \ (8\ \mathrm{TeV}) & &\\ \hline
\end{array}
$
\end{center}
\caption{Table of production cross-section times branching ratios over standard model values for CMS \cite{CMS:2012gu}, ATLAS \cite{Atlas:2012gk} and the Tevatron \cite{Tevatron:2012zzl}.}
\label{APP:tableofmus}\end{table}

The limit for $\Delta \rho$ reported in \cite{Aaltonen:2012bp,Abazov:2012bv,Group:2012gb,Barger:2012hr} is
\begin{align}
\Delta \rho = (4.2 \pm 2.7) \times 10^{-4}.
\end{align}

To get bounds from $b \to s \gamma$, we define the ratio of SUSY to SM contributions  \cite{Misiak:2004ew,Misiak:2006ab,Misiak:2006zs,Freitas:2008vh,Asner:2010qj,:2012iwb}
\begin{equation}
R \equiv \frac{\text{BR}(b \to s \gamma)_\text{SUSY}}{\text{BR}(b \to s \gamma)_\text{SM}} 
\end{equation}
Adding to the  uncertainty of the SM prediction ${\rm Br} (B \to X_s \gamma)_{\rm SM} = (3.15 \pm 0.23) \cdot 10^{-4}$ an intrinsic SUSY error of $0.15$ as well as the error of the experimental world average ${\rm Br} (B \to X_s \gamma)_{\rm exp} = (3.43 \pm 0.22) \cdot 10^{-4}$~\cite{Amhis:2012bh}, leads to the following $95\%$~CL bound 
\begin{equation}
R = [0.87, 1.31] \, .
\end{equation}

\end{document}

%% file: lambdas.tex
\begin{align}
128\pi^2 \lambda_1 =& 4 \lambda_{S}^{4} {\log\Big[\tilde{m}_{SI}^2 v^{-2} \Big]} +4 \lambda_{S}^{4} {\log\Big[\tilde{m}_{SR}^2 v^{-2} \Big]} +4 \lambda_{T}^{4} {\log\Big[\tilde{m}_{TI}^2 v^{-2} \Big]} +4 \lambda_{T}^{4} {\log\Big[\tilde{m}_{TR}^2 v^{-2} \Big]} \nonumber \\ 
 &+8 \Big(- \tilde{m}_{TI}^2  + \tilde{m}_{SI}^2\Big)^{-1} \Big(- \tilde{m}_{TR}^2  + \tilde{m}_{SR}^2\Big)^{-1} \lambda_{S}^{2} \lambda_{T}^{2} \Big(-2 \tilde{m}_{SI}^2 \tilde{m}_{SR}^2 +2 \tilde{m}_{SR}^2 \tilde{m}_{TI}^2 + \nonumber \\ 
 &2 \tilde{m}_{SI}^2 \tilde{m}_{TR}^2 -2 \tilde{m}_{TI}^2 \tilde{m}_{TR}^2 +\tilde{m}_{SI}^2 \Big(- \tilde{m}_{TR}^2  + \tilde{m}_{SR}^2\Big){\log\Big[\tilde{m}_{SI}^2 v^{-2} \Big]} \nonumber \\ 
 &+\tilde{m}_{SR}^2 \Big(- \tilde{m}_{TI}^2  + \tilde{m}_{SI}^2\Big){\log\Big[\tilde{m}_{SR}^2 v^{-2} \Big]} - \tilde{m}_{SR}^2 \tilde{m}_{TI}^2 {\log\Big[\tilde{m}_{TI}^2 v^{-2} \Big]} +\tilde{m}_{TI}^2 \tilde{m}_{TR}^2 {\log\Big[\tilde{m}_{TI}^2 v^{-2} \Big]} \nonumber \\ 
 &- \tilde{m}_{SI}^2 \tilde{m}_{TR}^2 {\log\Big[\tilde{m}_{TR}^2 v^{-2} \Big]} +\tilde{m}_{TI}^2 \tilde{m}_{TR}^2 {\log\Big[\tilde{m}_{TR}^2 v^{-2} \Big]} \Big)\nonumber \\ 
 &+4 \Big(- \tilde{m}_{TR}^2  + \tilde{m}_{TI}^2\Big)^{-1} \Big(- \Big(- \tilde{m}_{TR}^2  + \tilde{m}_{TI}^2\Big)\Big(-2 \lambda_{T}^{2}  + g_{2}^{2}\Big)^{2} \nonumber \\ 
 &+\Big(2 \Big(3 \tilde{m}_{TI}^2  - \tilde{m}_{TR}^2 \Big)\lambda_{T}^{4}  -4 g_{2}^{2} \tilde{m}_{TI}^2 \lambda_{T}^{2}  + g_{2}^{4} \tilde{m}_{TI}^2 \Big){\log\Big[\tilde{m}_{TI}^2 v^{-2} \Big]} \nonumber \\ 
 &+\Big(2 \Big(-3 \tilde{m}_{TR}^2  + \tilde{m}_{TI}^2\Big)\lambda_{T}^{4}  + 4 g_{2}^{2} \tilde{m}_{TR}^2 \lambda_{T}^{2}  - g_{2}^{4} \tilde{m}_{TR}^2 \Big){\log\Big[\tilde{m}_{TR}^2 v^{-2} \Big]} \Big) 
\end{align}
\begin{align}
128\pi^2\lambda_2 =& 4 \lambda_{S}^{4} {\log\Big[\tilde{m}_{SI}^2 v^{-2} \Big]} +4 \lambda_{S}^{4} {\log\Big[\tilde{m}_{SR}^2 v^{-2} \Big]} +4 \lambda_{T}^{4} {\log\Big[\tilde{m}_{TI}^2 v^{-2} \Big]} +4 \lambda_{T}^{4} {\log\Big[\tilde{m}_{TR}^2 v^{-2} \Big]} \nonumber \\ 
 &+8 \Big(- \tilde{m}_{TI}^2  + \tilde{m}_{SI}^2\Big)^{-1} \Big(- \tilde{m}_{TR}^2  + \tilde{m}_{SR}^2\Big)^{-1} \lambda_{S}^{2} \lambda_{T}^{2} \Big(-2 \tilde{m}_{SI}^2 \tilde{m}_{SR}^2 +2 \tilde{m}_{SR}^2 \tilde{m}_{TI}^2 +2 \tilde{m}_{SI}^2 \tilde{m}_{TR}^2 \nonumber \\ 
 & -2 \tilde{m}_{TI}^2 \tilde{m}_{TR}^2 +\tilde{m}_{SI}^2 \Big(- \tilde{m}_{TR}^2  + \tilde{m}_{SR}^2\Big){\log\Big[\tilde{m}_{SI}^2 v^{-2} \Big]} \nonumber \\ 
 &+\tilde{m}_{SR}^2 \Big(- \tilde{m}_{TI}^2  + \tilde{m}_{SI}^2\Big){\log\Big[\tilde{m}_{SR}^2 v^{-2} \Big]} - \tilde{m}_{SR}^2 \tilde{m}_{TI}^2 {\log\Big[\tilde{m}_{TI}^2 v^{-2} \Big]} +\tilde{m}_{TI}^2 \tilde{m}_{TR}^2 {\log\Big[\tilde{m}_{TI}^2 v^{-2} \Big]} \nonumber \\ 
 &- \tilde{m}_{SI}^2 \tilde{m}_{TR}^2 {\log\Big[\tilde{m}_{TR}^2 v^{-2} \Big]} +\tilde{m}_{TI}^2 \tilde{m}_{TR}^2 {\log\Big[\tilde{m}_{TR}^2 v^{-2} \Big]} \Big)\nonumber \\ 
 &+4 \Big(- \tilde{m}_{TR}^2  + \tilde{m}_{TI}^2\Big)^{-1} \Big(- \Big(- \tilde{m}_{TR}^2  + \tilde{m}_{TI}^2\Big)\Big(-2 \lambda_{T}^{2}  + g_{2}^{2}\Big)^{2} \nonumber \\ 
 &+\Big(2 \Big(3 \tilde{m}_{TI}^2  - \tilde{m}_{TR}^2 \Big)\lambda_{T}^{4}  -4 g_{2}^{2} \tilde{m}_{TI}^2 \lambda_{T}^{2}  + g_{2}^{4} \tilde{m}_{TI}^2 \Big){\log\Big[\tilde{m}_{TI}^2 v^{-2} \Big]} \nonumber \\ 
 &+\Big(2 \Big(-3 \tilde{m}_{TR}^2  + \tilde{m}_{TI}^2\Big)\lambda_{T}^{4}  + 4 g_{2}^{2} \tilde{m}_{TR}^2 \lambda_{T}^{2}  - g_{2}^{4} \tilde{m}_{TR}^2 \Big){\log\Big[\tilde{m}_{TR}^2 v^{-2} \Big]} \Big)
\end{align}
\begin{align}
64\pi^2\lambda_3^\prime =& 2 \lambda_{S}^{4} {\log\Big[\tilde{m}_{SI}^2 v^{-2} \Big]} +2 \lambda_{S}^{4} {\log\Big[\tilde{m}_{SR}^2 v^{-2} \Big]} +2 \lambda_{T}^{4} {\log\Big[\tilde{m}_{TI}^2 v^{-2} \Big]} +2 \lambda_{T}^{4} {\log\Big[\tilde{m}_{TR}^2 v^{-2} \Big]} \nonumber \\ 
 &-4 \Big(- \tilde{m}_{TI}^2  + \tilde{m}_{SI}^2\Big)^{-1} \Big(- \tilde{m}_{TR}^2  + \tilde{m}_{SR}^2\Big)^{-1} \lambda_{S}^{2} \lambda_{T}^{2} \Big(-2 \tilde{m}_{SI}^2 \tilde{m}_{SR}^2 +2 \tilde{m}_{SR}^2 \tilde{m}_{TI}^2 \nonumber \\ 
 &+2 \tilde{m}_{SI}^2 \tilde{m}_{TR}^2 -2 \tilde{m}_{TI}^2 \tilde{m}_{TR}^2 +\tilde{m}_{SI}^2 \Big(- \tilde{m}_{TR}^2  + \tilde{m}_{SR}^2\Big){\log\Big[\tilde{m}_{SI}^2 v^{-2} \Big]} \nonumber \\ 
 &+\tilde{m}_{SR}^2 \Big(- \tilde{m}_{TI}^2  + \tilde{m}_{SI}^2\Big){\log\Big[\tilde{m}_{SR}^2 v^{-2} \Big]} - \tilde{m}_{SR}^2 \tilde{m}_{TI}^2 {\log\Big[\tilde{m}_{TI}^2 v^{-2} \Big]} +\tilde{m}_{TI}^2 \tilde{m}_{TR}^2 {\log\Big[\tilde{m}_{TI}^2 v^{-2} \Big]} \nonumber \\ 
 &- \tilde{m}_{SI}^2 \tilde{m}_{TR}^2 {\log\Big[\tilde{m}_{TR}^2 v^{-2} \Big]} +\tilde{m}_{TI}^2 \tilde{m}_{TR}^2 {\log\Big[\tilde{m}_{TR}^2 v^{-2} \Big]} \Big)\nonumber \\ 
 &+2 \Big(- \tilde{m}_{TR}^2  + \tilde{m}_{TI}^2\Big)^{-1} \Big(- \Big(- \tilde{m}_{TR}^2  + \tilde{m}_{TI}^2\Big)\Big(-2 \lambda_{T}^{2}  + g_{2}^{2}\Big)^{2} \nonumber \\ 
 &+\Big(2 \Big(3 \tilde{m}_{TI}^2  - \tilde{m}_{TR}^2 \Big)\lambda_{T}^{4}  -4 g_{2}^{2} \tilde{m}_{TI}^2 \lambda_{T}^{2}  + g_{2}^{4} \tilde{m}_{TI}^2 \Big){\log\Big[\tilde{m}_{TI}^2 v^{-2} \Big]} \nonumber \\ 
 &+\Big(2 \Big(-3 \tilde{m}_{TR}^2  + \tilde{m}_{TI}^2\Big)\lambda_{T}^{4}  + 4 g_{2}^{2} \tilde{m}_{TR}^2 \lambda_{T}^{2}  - g_{2}^{4} \tilde{m}_{TR}^2 \Big){\log\Big[\tilde{m}_{TR}^2 v^{-2} \Big]} \Big)
\end{align}